\documentstyle[12pt,aaspp4,eqsecnum,flushrt]{article}
\def\kms{km~s$^{-1}$}
\begin{document}
\title{Lyman-Alpha Absorption Systems and\\ 
the Nearby Galaxy Distribution}

\author{Norman A. Grogin and Margaret J. Geller} 

\affil{Harvard-Smithsonian Center for Astrophysics,
60 Garden Street, Cambridge, MA 02138\\ 
E-mail: ngrogin,mgeller@cfa.harvard.edu} 

\slugcomment{Submitted to the {\it Astrophysical Journal}} 

\abstract{ 
We study the galaxy number density (smoothed on a $5h^{-1}$~Mpc scale)
around 18 low-redshift Lyman-alpha absorbers previously observed with
{\sl HST}.  The absorbers lie in the foregrounds of Mrk 335, Mrk 421,
Mrk 501, I Zw 1, and 3C 273, all within regions where there are now
complete redshift surveys to $m_{\rm Zw}=15.5$.  We construct a
smoothed galaxy number density field from the redshift survey data and
determine the distribution of densities at the Lyman-alpha absorber
locations.  We also find the distribution of galaxy number density for
a variety of test samples: all galaxy locations within the Center for
Astrophysics Redshift Survey (CfA2), CfA2 galaxy locations along
randomly selected lines of sight, and randomly chosen redshifts along
random lines of sight.

The Lyman-alpha absorbers are present in dense regions of the survey,
but occur far more frequently in underdense regions than do typical
luminous galaxies.  The distribution of smoothed galaxy density around
the Lyman-alpha absorbers is inconsistent at the 4$\sigma$ level with
the density distribution around survey galaxies.  It is highly
consistent with a density distribution at randomly chosen redshifts
along random lines of sight.  This supports earlier evidence that the
nearby, low column density ($\log N_{H\sc i} \lesssim 14$) Lyman-alpha
forest systems are spatially distributed at random; they are not well
correlated with the local large-scale structure.
}


\begin{section}{Introduction}
The spectra of high-redshift QSOs display a dense network of
absorption lines blueward of the QSO Ly$\alpha$ emission, first
recognized by Lynds (1971) as low column density Ly$\alpha$ absorption
in the rest frame of intervening discrete systems (cf.~reviews by
\cite{bat93}; \cite{wey93}).  With the advent of the {\sl HST}, this
``Ly$\alpha$ forest'' could be studied at wavelengths below the
atmospheric UV cutoff, corresponding to absorbers with redshift
$z\lesssim1.6$.  Because the Ly$\alpha$ forest reveals a considerable
population of very tenuous structures ($N_{H\sc i} >
10^{12}$~cm$^{-2}$) across cosmological distances ($z \lesssim 4$),
there has been great interest in determining the nature of the
absorbers: galactic, intergalactic, or both.

Early studies of the redshift distribution of Ly$\alpha$ forest clouds
at high $z$ revealed that the metal-poor population was much less
clustered than the present-day galaxies (\cite{sarg80}).  Furthermore,
there were suggestions that the Ly$\alpha$ forest might be an entirely
separate population from the absorbers at larger column density
($N_{H\sc i} \geq 10^{17}$~cm$^{-2}$), which have associated
metal-line absorption and cluster like galaxies (\cite{sarg88}; \cite{tyt87}).
Proposed explanations of a non-galactic Ly$\alpha$ forest include
pressure-confined clouds in the IGM (\cite{sarg80}; \cite{iktu91}) and CDM
minihalos (\cite{ik86}; \cite{rees86}).  Recent results from the HIRES
spectrograph on the Keck 10m telescope still admit the possibility of
a dual-population (galactic and intergalactic) Ly$\alpha$ forest.
Highly resolved metal-line absorption in high-$z$ damped-Ly$\alpha$
systems appears to originate in thick, rapidly rotating ($\gtrsim
200$~\kms) protogalactic disks (\cite{pw97}).  The mean metallicity of
high-$z$ Ly$\alpha$ clouds declines sharply at low column densities,
from [C/H]$\simeq -2.5$ at $N_{H\sc i} \geq 10^{14.5}$~cm$^{-2}$ to an
upper limit of [C/H]$< -3.5$ for $10^{13.5} < N_{H\sc i} <
10^{14}$~cm$^{-2}$ (\cite{lu98}).

With the {\sl HST} discovery of several $z\leq0.16$ Ly$\alpha$
clouds along the 3C 273 sightline (\cite{mor91}; \cite{b91}), it was now
possible with pencil-beam redshift surveys to search for
far less luminous galaxies coincident in redshift space with the low
column density absorbers.  Morris et al.~(1993)
completed a $B\sim19$ survey of $\sim\!4$ square degrees around 3C 273
that detected no galaxies within $1.5h^{-1}$~Mpc of any absorber.
They concluded that the absorption is extragalactic
in origin.  A further deep imaging survey of $53'\times52'$ around 3C 273 by
Rauch, Weymann, \& Morris (1996) to a limiting central surface
brightness of $\mu_r=26.4$ revealed no low surface brightness
galaxies within projected separations of $80h^{-1}$~kpc and
$123h^{-1}$~kpc, respectively, at the two nearest absorber redshifts.

Lanzetta et al.~(1995), on the basis of an $r<21.5$ survey within
$1\farcm3$ of several {\sl HST}/FOS target QSOs (\cite{bah93}),
concluded instead that the Ly$\alpha$ forest is largely due to
extended galactic gaseous halos (\cite{bs69}) or gaseous disks
(\cite{mal93}; \cite{hoff93}) of size $\approx160h^{-1}$~kpc and roughly unit
covering factor.  Their survey spanned a much greater cumulative
pathlength, $\Delta z\approx2.86$, than the $\Delta z\approx0.16$
sampled by the Morris et al.~(1993) study of the 3C 273 foreground.
Lanzetta et al.\@ detected 30 galaxies at impact parameters
$\lesssim160h^{-1}$~kpc from the QSO lines of sight (LOS), closer than
any detected galaxy to the 3C 273 LOS.  They found: (1) a strong
increase in the probability of a galaxy-absorber association for
galaxy impact parameters under $160h^{-1}$~kpc.  They also noted a
statistical anti-correlation between the equivalent width of
Ly$\alpha$ absorption and the impact parameter of a galaxy-absorber
pair.  They argue that at $z\lesssim1$, the fraction of Ly$\alpha$
absorption systems arising in galactic halos is at least
$0.35\pm0.10$, and probably $0.65\pm0.18$.  It is noteworthy that
their {\sl HST}/FOS Ly$\alpha$ absorber sample was limited to
equivalent widths $W_{\rm Ly\alpha} \gtrsim 0.3$\AA, larger than
almost all the {\sl HST}/GHRS detections and probably not
representative of the extremely metal-poor population, with $N_{H\sc
i} < 10^{14}$~cm$^{-2}$.  Mo and Morris (1994) used Monte Carlo
simulations to show that the galactic-halo fraction of 3C 273
absorbers, all with $N_{H\sc i}\lesssim10^{14}$~cm$^{-2}$, could only
be $\approx20$\% given the observed number of absorbers and the weak
correlation between absorbers and galaxies on large scales.

In a different approach, Shull, Stocke, \& Penton (1996, hereafter
SSP96; \cite{ssp95}) obtained {\sl HST}/GHRS spectra of four nearby
bright AGNs within the boundaries of existing wide-angle redshift
surveys to determine whether Ly$\alpha$ absorbers nearby are
correlated with large-scale structure, as first suggested by Oort
(1981).  They detected a total of 11 Ly$\alpha$ absorbers along these
sightlines within a GHRS wavelength range corresponding to
$1500\lesssim cz \lesssim 10500$~\kms.  Using the wavelet analysis of
Slezak, de Lapparent, \& Bijaoui (1993) to subdivide their AGN
sightlines into ``void'' and ``supercluster'' regions, SSP96 found
four of their eleven absorbers in voids.  Because their sightlines probed
an approximately equal pathlength of void and supercluster, binomial
statistics gave a fair probability (19\%) that the $z<0.035$ absorbers
are uniformly distributed and not biased towards superclusters.  From
the relation between Ly$\alpha$ equivalent widths and nearest-galaxy
distance, SSP96 argue that the $W_{\rm Ly\alpha}< 0.1$\AA\
absorbers are distributed in a manner statistically indistinguishable
from clouds randomly placed with respect to galaxies.  Although the
nearest-galaxy distances of their $W_{\rm Ly\alpha}> 0.1$\AA\ subsample
are broadly consistent with the Lanzetta et al.~(1995) trend extended
to larger impact parameters ($\gtrsim500h^{-1}$~kpc), it is unlikely
that the Ly$\alpha$ absorbers could be physically associated with
galaxies at that separation.

We use the large-scale smoothed galaxy number density to compare the
distribution of $z<0.035$ Ly$\alpha$ absorbers with that of typical luminous
galaxies.  Whereas SSP96 identified the absorbers as either ``void'' or
``supercluster'', we quantify the large-scale structure in the
vicinity of the absorbers.  The smoothed galaxy density field is a
more robust indicator of absorber environment than the nearest-galaxy
distance, particularly on scales larger than 1~Mpc.  

The technique we develop requires a wide-angle redshift survey
overlapping the AGN sightlines of interest and adequately sampling the
galaxy population at the Ly$\alpha$ absorber redshifts.  Thus we
restrict ourselves to the $cz < 10500$~\kms\ absorbers along the
sightlines to 3C 273 and the four AGNs of SSP96.  For the 18
Ly$\alpha$ absorbers meeting these specifications, we use the galaxy
number density field from the CfA2 Redshift Survey
(\cite{gh89}; \cite{v94}; \cite{mhg94}, hereafter MHG94) and adjacent
regions also surveyed to $m_{Zw} = 15.5$ (\cite{marzvel}; \cite{3cz}).  Even
with this modest sample of absorbers, we obtain rather tight
constraints on the relation between the local Ly$\alpha$ absorbers and
the large-scale structure marked by intrinsically luminous galaxies.

Section \ref{datasec} describes the nearby Ly$\alpha$ absorption
systems and the galaxy redshift surveys surrounding them.  Section
\ref{densec} describes the density estimation technique and shows maps
of the galaxy density field toward the five target AGNs.  We also
derive the cumulative distribution function (CDF) of the densities
around the 18 Ly$\alpha$ absorption systems.  In \S\ref{results} we
investigate the implications of the Ly$\alpha$ density CDF by
comparing it with CDFs for several test populations.  We also compute
likelihood intervals for models of the absorbers' spatial
distribution.  We address sources of uncertainty in \S\ref{discsec},
and estimate the improved constraints likely with a larger sample of
nearby Ly$\alpha$ absorbers.  We conclude in \S\ref{concsec}.
\end{section}
\begin{section}{Data} \label{datasec}
\begin{subsection}{Lyman-alpha Absorption Systems}
Our sample of local Ly$\alpha$ absorbers comprises seven systems
with $cz < 10500$ \kms\ detected towards 3C 273 (\cite{mor93}) and eleven
systems detected in the sightlines toward the AGN Mrk 335, Mrk 421, Mrk
501, and I Zw 1 (\cite{ssp96}).  The 3C 273 observations extend down to 1215\AA\
(\cite{wey95}) and include Ly$\alpha$ forest systems well
into the damping wing of Galactic Ly$\alpha$ absorption.  The AGN
observations span a more limited wavelength range of 1222\AA--1259\AA,
corresponding to a redshift window of $1500 \lesssim cz \lesssim 
10500$~\kms\ for the detection of Ly$\alpha$ absorption systems.  Three
of the four AGN have $cz < 10500$~\kms, further
reducing the path length probed by these sightlines.  Table
\ref{srctable} summarizes the relevant data for each of the
Ly$\alpha$ sources, including the source redshift and the redshift range
for absorber detection.  In Table \ref{abstable} we list the redshifts
and equivalent widths for the 18 Ly$\alpha$ absorbers.
\placetable{srctable}
\placetable{abstable}
\end{subsection}
\begin{subsection}{Redshift Surveys} \label{czdatasec}
The CfA2 Redshift Survey (\cite{gh89}) provides the broad sky coverage
and dense sampling necessary to evaluate the galaxy number density,
smoothed on scales $\lesssim5h^{-1}$~Mpc, around Ly$\alpha$ absorbers
with $cz \lesssim10000$~\kms.  Although there have been much deeper,
pencil-beam surveys along QSO sightlines (\cite{mor93}; \cite{l95}) to
link Ly$\alpha$ absorption with specific galaxies, these subtend much
too small an angle to evaluate nearby densities on $5h^{-1}$~Mpc
scales.  SSP96 chose their four AGN partly because they are within the
CfA2 boundaries as analyzed by Vogeley et al.~(1994) and MHG94:
$8^{\rm h} \leq \alpha \leq 17^{\rm h}$, $8\fdg5 \leq \delta \leq
44\fdg5$ (CfA2 North) and $20^{\rm h} \leq \alpha \leq 4^{\rm h}$,
$-2\fdg5 \leq \delta \leq 42\arcdeg$ (CfA2 South).  Redshifts around
these four AGN sightlines are contained in Huchra et al.~(1990),
Huchra, Geller, \& Corwin (1995), Huchra, Vogeley, \& Geller (1998),
and are available through NED\footnote{The NASA/IPAC Extragalactic Database
(NED) is operated by the Jet Propulsion Laboratory, California
Institute of Technology, under contract with the National Aeronautics
and Space Administration.}.

We have also surveyed a $2^{\rm
h}\times12\arcdeg$ region nearly centered on 3C 273 (\cite{3cz}).
With $\approx1060$ redshifts in the region $11^{\rm h}30^{\rm m} \leq
\alpha \leq 13^{\rm h}30^{\rm m}$ and $-3\fdg5 \leq \delta \leq
8\fdg5$, this survey extension is 99.5\% complete to the CfA2
magnitude limit ($m_{\rm Zw} = 15.5$) and 98.1\% complete to the
Zwicky catalog limit ($m_{\rm Zw} = 15.7$).  This additional survey
yields estimates of the global density at the locations of the 3C 273
Ly$\alpha$ absorbers.  In addition to the 3C 273 field redshifts, we
also incorporate $\sim\!1700$ Zwicky galaxy redshifts at low Galactic
latitude from the Galactic Plane Survey by Marzke, Geller, \& Huchra
(1996).  These additional redshifts outside the CfA2 boundaries
prevent substantial underestimation of the galaxy density at locations
within a smoothing length of the survey boundaries.
\end{subsection}
\end{section}

\begin{section}{Estimating the Smoothed Galaxy Number Density} \label{densec}
Our goal is to compare the distribution of Ly$\alpha$ absorbers with
the large-scale distribution of galaxies.  If the absorbers sample the
large-scale structure marked by luminous galaxies in a redshift
survey, the distribution of smoothed galaxy number densities around
the absorbers should be consistent with the distribution of densities
around survey galaxies.  Conversely, if the Ly$\alpha$ absorbers arise
in a host population distributed independently of the large-scale
structure, the densities around the absorbers should resemble the
densities around similarly distributed locations in the survey volume.

In \S\ref{denmethsec} we describe the smoothed galaxy number density
field constructed from the redshift survey data of \S\ref{czdatasec}.
We show maps of the density field towards five AGN with low-redshift
Ly$\alpha$ absorbers in \S\ref{sliceplots}.  Finally we combine the
densities at each of the 18 absorber locations into a cumulative
distribution function (\S\ref{lyalcdfsec}) for comparison with density
CDFs of various test populations in \S\ref{results}.

\begin{subsection}{Method} \label{denmethsec}
To evaluate the galaxy density around the local Ly$\alpha$ absorbers,
we first transform the point distribution of the CfA2 Survey into a
continously defined number density field throughout redshift
space.  We smooth each galaxy in redshift space by a unit-normalized
Gaussian kernel $W$ of width $\sigma = 5h^{-1}$~Mpc:
\begin{equation}
W\left({{\bf x}-{\bf x}_{\rm gal}}\right) = \left({2\pi \sigma}\right)^{-3/2} 
	\exp\left({\left|{{\bf x}-{\bf x}_{\rm gal}}\right|^2/2\sigma^2}\right).
\end{equation}
We choose a $5h^{-1}$~Mpc smoothing length to coincide
with the galaxy-galaxy correlation length (\cite{peeb}; \cite{pecvel}; 
\cite{lcrscorr}) and with the pairwise velocity dispersion in the
survey (\cite{pecvel}).  In \S\ref{smoothdisc}
we discuss the sensitivity of our results to this choice.

We make no attempt to remove peculiar velocity distortions (cluster
``fingers'', etc.) from the redshift survey or from the Ly$\alpha$
absorber velocities.  In principle, they should share the same local velocity
field.  We do introduce the standard heliocentric to
Galactocentric correction,
\begin{equation}
cz = cz_{\sun} + (300\,\hbox{\rm km s}^{-1}) \sin l \cos b,
\end{equation}
for an object at Galactic longitude $l$ and latitude $b$.  We place
each object at a comoving distance $r$ appropriate for a $q_0 = 0.5$
universe with pure Hubble flow:
\begin{equation} \label{comove}
r(z) = \left({{2c}\over{H_0}}\right) \left[{1 - (1 + z)^{-1/2}}\right].  
\end{equation}
We thus underestimate spatial overdensities associated with clusters,
which are broadened in the radial direction.  Our smoothing kernel
effectively washes out peculiar velocities $\lesssim 500$~\kms, close
to the $540\pm180$~\kms\ pairwise velocity dispersion measured by
Marzke et al.~(1995) for the combined CfA2 and SSRS2 (\cite{ssrs2})
surveys.

Because the CfA2 Redshift Survey is flux-limited, an increasing fraction
of the galaxies at larger redshift fall below the magnitude limit and do
not appear in the survey.  In computing the density field, we compensate 
for the magnitude-limited sample by assigning each galaxy
a weight $1/\psi$, where the selection function $\psi$ is
\begin{equation} \label{selfneq}
\psi(\alpha,\delta,z) = {{\displaystyle\int_{-\infty}^{M_{\rm lim}(\alpha,\delta,z)}
{\phi(M)\,dM}}
\over{\displaystyle\int_{-\infty}^{M_{\rm cut}}{\phi(M)\,dM}}}.
\end{equation}
Here $M_{\rm lim}$ is the effective absolute magnitude limit at the
galaxy position and $\phi(M)$ is the differential luminosity function
(LF).  For $M_{\rm lim}\geq M_{\rm cut}$, we assign galaxies unit weight.  All
magnitudes are Zwicky magnitudes. Numerical values for absolute
magnitudes implicitly include the $h$-dependence in equation
(\ref{comove}).  

MHG94 fit the CfA2 LF to a Schechter function
$\phi_{\rm SF}$ (\cite{sch76}), convolved with a Gaussian error of
$\sigma_M = 0.35$~mag (\cite{h76}) in the Zwicky magnitudes:
\begin{eqnarray}
\phi_{\rm SF}(M) &=& \phi_* \, (0.4 \ln 10) \, 10^{0.4\, (M_* - M)\, (1 + \alpha)} 
	\exp\left[{-10^{0.4 (M_* - M)}}\right]; \nonumber \\
\phi_{\rm CfA2}(M) &=& \label{cfa2lfeq}
{1\over{\sqrt{2\pi}\sigma_M}} \int_{-\infty}^\infty{\phi_{\rm SF}(M')
\exp\left[{-(M'-M)^2/2\sigma_M^2)}\right]\,dM'}.
\end{eqnarray}
We adopt the values $\phi_* = 0.04\,({\rm Mpc}/h)^{-3}$, $M_* =
-18.8$, and $\alpha = -1.0$ found by MHG94 using $M_{\rm cut}=-16.5$.
We discuss the sensitivity of our results to the LF parameters
in \S\ref{lfdisc}.  For computational convenience in determining
$\psi$, we replace the convolution of equation (\ref{cfa2lfeq}) with
$\phi_{\rm CfA2}(M) \approx \phi_{\rm SF}(M + 0.1\,{\rm mag})$.  This
approximation recovers the true $\psi$ to better than 5\% for
$cz\lesssim12000$~\kms.

For a galaxy at position ($\alpha,\delta$) and at luminosity distance $D_L(z) = 
(1+z)\,r(z)$, we estimate $M_{\rm lim}$ according to
\begin{equation}
M_{\rm lim}(\alpha,\delta,z) = m_{\rm lim} - 5 \log\left[{{(1+z)\,r(z)}\over{1h^{-1}\,
\rm{Mpc}}}\right] - 25 - \Delta m_{\rm K}(z)  
+ \Delta m_{\rm ext}(\alpha,\delta). \label{mlimeq}
\end{equation}
In equation (\ref{mlimeq}), $m_{\rm lim}$ is the CfA2 flux limit,
$\Delta m_{\rm K}$ is a $K$-correction, and $\Delta m_{\rm ext}$ is a
correction for Galactic extinction.  Photoelectric photometry of
Zwicky galaxies (\cite{zwsc}) suggests that Volume I of the Zwicky
catalog (\cite{zw61}) goes $\approx0.4$~mag fainter than the other
volumes at the CfA2 magnitude limit, $m_{\rm Zw} = 15.5$.  To correct
for this Volume I scale error, we adopt $m_{\rm lim}=15.9$ for all
galaxies with $\delta
\leq 14\fdg5$ in CfA2 North and the 3C 273 region extension.

Lacking morphological types for the majority of CfA2, we
apply a generic $K$-correction, $\Delta m_{\rm K}(z) = 3z$, appropriate
for type Sab (\cite{kcorr}).  Our error in the K-correction will be
small due to the predominantly low redshifts ($z \lesssim 0.05$) in
the survey.  To obtain our correction $\Delta m_{\rm ext}(\alpha,\delta)$
for Galactic extinction along a particular line of sight, we
first interpolate the H{\sc i} map of Stark et al.~(1992).  We
then convert from H{\sc i} to reddening with the relation
$\left<{N({\rm H}{\sc i})/E(\bv)} \right> = 4.8 \times 10^{21}$
cm$^{-2}$~mag$^{-1}$ (\cite{zom90}), and adopt a standard extinction
law $\Delta m_{\rm ext} \equiv A_B = 4.0 E(\bv)$.

We compute the smoothed galaxy number density $n$ at a
given point ${\bf r} \equiv (\alpha,\delta,r(z))$ by summing the 
contributions from all $i$ galaxies in the survey:
\begin{eqnarray}
n({\bf r}) &=& \sum_i{{W({\bf r} - {\bf r}_i)}\over\psi_i}  \nonumber \\
	&=& \sum_i{{W({\bf r} - {\bf r}_i) 
\displaystyle\int_{-\infty}^{M_{\rm cut}}{\phi(M)\,dM}
}\over{\displaystyle\int_{-\infty}^{M_{\rm lim}({\bf r}_i)}
{ \phi(M)\,dM}}} \nonumber \\
&=& \bar n \sum_i{{W({\bf r} - {\bf r}_i) 
\over{\displaystyle\int_{-\infty}^{M_{\rm lim}({\bf r}_i)}{ \phi(M)\,dM}}}}.
\end{eqnarray}
For the CfA2 Survey, MHG94 derive a mean density $\bar n \equiv
\int_{-\infty}^{M_{\rm cut}}{\phi(M)\,dM} = 0.07\,({\rm Mpc}/h)^{-3}$
with $M_{\rm cut}=-16.5$.  This analysis yields a distribution of
dimensionless galaxy density contrasts, $n({\bf r}_i)/\bar n$, at the
absorber positions ${\bf r}_i$ which may be compared with the
$n/\bar n$ distribution of various control samples.
\end{subsection}
\begin{subsection}{CfA2 Density Maps toward Local Lyman-alpha Systems}
\label{sliceplots}
For each of the five sources in Table \ref{srctable}, we display the
results of our density mapping in two forms .  The first
representation is similar to the standard ``wedge plot'' format for
redshift survey data: we project galaxy positions over a narrow
declination range onto a circular section with R.A. as the azimuth
coordinate and redshift as the radial coordinate.  We add the density
contrast as a superposed isodensity contour map, made by sampling
the density at regular $3h^{-1}$~Mpc intervals on the midplane (or,
more accurately, the mid-cone) of the wedge plot.

We set the outer boundary of the wedge plots at 15000~\kms, well into
the region where shot noise dominates the density estimate.  We choose a
``thickness'' of six degrees in declination, centered on each LOS.  Our
$5h^{-1}$~Mpc density smoothing length subtends this angle at a
redshift of $\approx5000$~\kms.  For $cz\gtrsim5000$~\kms, there is density
variation with declination that is lost in projection.  To showcase the
large-scale structure toward each target, the wedge plots have a
30\arcdeg\ opening angle centered on each LOS, corresponding to ($2^{\rm
h}/\cos\delta$) of right ascension at the central declination $\delta$.
Our density smoothing length subtends this angle at a redshift of
($1000/\cos\delta$)~\kms.  We accommodate the dynamic range of the
density map with underdensity contours (dotted) in linear decrements
of $0.2\bar n$ and overdensity contours (solid) in logarithmic
increments corresponding to $\bar n$, $2\bar n$, $4\bar n$, $8\bar n$,
etc.  In addition to the surveyed galaxies (small squares), we also
plot the locations of the detected Ly$\alpha$ absorbers (large
triangles) along the line of sight (dashed) to the background source
(large circle, if $cz_{\rm src} \leq 15000$~\kms).
	
For the second display, we sample the density at $100$~\kms\
radial intervals for $cz\leq15000$~\kms\ along the line of sight to
the selected background source.  We then plot the density profile
$(n/\bar n)$ along the LOS (dashed curve) versus redshift.  We
indicate the redshifts of the detected Ly$\alpha$ absorbers (dotted
lines) as well as the redshift interval over which the absorbers could
have been detected (solid lines).  If
the upper redshift limit coincides with the source of Ly$\alpha$
emission, we indicate this coincidence with a doubled solid line.
Within this redshift interval, we
also show the $\pm1\sigma$ uncertainty in the local density estimate (solid
curves).  This density error estimate is the quadrature sum of the fractional
errors due to sampling variance (cf.~\S\ref{shotsec}) and the uncertain
survey LF (cf.~\S\ref{lfdisc}).

\begin{subsubsection}{3C 273}
For the wedge plot of the 3C 273 region (Figure \ref{3cwedge}), we
center the right ascension on the 11\fh5--13\fh5 range of the CfA2
survey extension (\cite{3cz}) instead of the $\pm15\arcdeg$ about 3C
273 ($\alpha = 12^{\rm h}26\fm6$).  Because the survey extension is
complete to the Zwicky catalog limit, $m_{\rm Zw} = 15.7$, we also
show the locations of the fainter galaxies (small crosses) to compare
with the Ly$\alpha$ systems.  We use only the brighter $m_{\rm Zw}
\leq 15.5$ galaxies to derive the density estimate.  The
uncertainty in the density estimator rises dramatically for
$cz\gtrsim12000$~\kms (cf.~\ref{shotsec}), and exceeds $\pm1$~dex at
the two absorption systems near 15000~\kms.  We therefore only
consider the 3C 273 absorbers with $cz\leq 10500$~\kms, coincident
with the velocity maximum for the other AGN lines of sight.
\placefigure{3cwedge}

The seven 3C 273 absorbers with $cz\leq10500$~\kms\ inhabit a wide
range of environments.  The two nearest Ly$\alpha$ systems are near
the Virgo cluster, and thus are difficult to identify in the crowded
wedge plot.  Their redshifts are clearly indicated in the associated
density profile.  These two systems inhabit some of the
highest-density regions in the LOS.  In contrast, two of the other
absorbers inhabit the low-density extreme of the LOS, a large void at
$cz\approx9500$~\kms.
\end{subsubsection}
\begin{subsubsection}{Mrk 335}
Figure \ref{335wedge} shows the Mrk 335 region.  The
local galaxy densities at the Ly$\alpha$ absorber locations span the density 
range in the LOS redshift window, including a pair with $cz\approx2350$~\kms\ 
near the density minimum.
\placefigure{335wedge}
\end{subsubsection}
\begin{subsubsection}{Mrk 421}
Figure \ref{421wedge} shows the Mrk 421 region.  This LOS is
intriguing in that only one absorber appears within the
$\approx7500$~\kms\ sensitivity window; it resides in a
relatively underdense environment.
\placefigure{421wedge}
\end{subsubsection}
\begin{subsubsection}{Mrk 501}
The proximity of Mrk 501 ($\alpha = 16^{\rm h}52\fm2$) to the CfA2
North boundary at $17^{\rm h}$ poses an added challenge to
determination of the densities surrounding its Ly$\alpha$ absorbers.
Fortunately there is an essentially complete redshift survey
(\cite{marzvel}) beyond this boundary, mitigating edge effects on the
density estimate.  In the region $17^{\rm h} \leq \alpha \leq 18^{\rm
h}$ and $+34\arcdeg
\leq \delta \leq +46\arcdeg$, there are redshifts for 140 of the 150
Zwicky galaxies with $m_{\rm Zw} \leq 15.5$.

Rather than cropping the Mrk 501 wedge plot at the CfA2 survey
boundary, we set the upper R.A. limit in Figure \ref{501wedge} at
$18^{\rm h}$ in order to show the large-scale structure east of the
LOS.  As noted above, there are redshifts for more than 90\% of the
galaxies in the plotted region with $\alpha>17^{\rm h}$.  We therefore
have high confidence in the density estimates along the Mrk 501 LOS.
\placefigure{501wedge}

Interestingly, there are no absorbers in the region of large
overdensity beyond $cz\simeq8000$~\kms\ associated with the Hercules
supercluster.  On the contrary, two of the absorbers are near the LOS
density minimum, in a foreground void to the supercluster.  A third is
on the outskirts of this void, in a region of middling density.
\end{subsubsection}
\begin{subsubsection}{I Zw 1}
Figure \ref{izw1wedge} shows the I Zw 1 region.  The two nearest
absorption systems are located in a relatively shallow void; the
third system appears to be associated with the galaxy wall behind it
at $cz\approx5500$~\kms.  Beyond this wall, the LOS passes through a 
larger and emptier void containing no detected Ly$\alpha$ systems.
\placefigure{izw1wedge}
\end{subsubsection}
\end{subsection}
\begin{subsection}{Density Distribution for the Lyman-Alpha Absorber Sample}
\label{lyalcdfsec}
Figure \ref{allcfacdf} shows the cumulative distribution function (CDF)
of the densities surrounding the 18 Ly$\alpha$ absorbers (solid).  We
plot the density contrast $(n/\bar n)$ on a logarithmic axis.  Error
in the overall normalization of $(n/\bar n)$ translates all CDFs along
the x-axis but does not change their shape.  The distribution of
$(n/\bar n)$ for the Ly$\alpha$ absorbers is broad, with fully half
the absorbers in substantially underdense regions of the survey:
$(n/\bar n)< 0.4$.  This considerable fraction of low-$(n/\bar n)$
absorption systems is the principal discriminant among the density
CDFs of test samples in \S\ref{results}.
\placefigure{allcfacdf}
\end{subsection}
\end{section}

\begin{section}{Interpretation} \label{results}
To understand the physical implications of the galaxy density
distribution around the sample of nearby Ly$\alpha$ absorption
systems, we define several test populations.  We then calculate the
local densities surrounding each of the locations in the test samples.
Finally, we employ a two-population Kolmogorov-Smirnov
(K-S) statistic to compare the galaxy number density distribution of the
Ly$\alpha$ absorbers with the density distributions from the test
samples of: (1) CfA2 galaxies (\S\ref{allcfa2desc},
\S\ref{loscfa2desc}) and (2) randomly selected locations within the
redshift survey boundaries (\S\ref{maxlikedesc}).  To construct these
test samples, we must also simulate the radial sampling bias caused by
the redshift windows of the Ly$\alpha$ forest observations (Table
\ref{srctable}).
\begin{subsection}{Comparison with All CfA2 Galaxies} \label{allcfa2desc}
For this test sample, we determine the density contrast $(n/\bar n)$ at the
locations of each CfA2 galaxy (within the boundaries listed in 
\S\ref{czdatasec} and including the 3C 273-region extension).
To account for the radial sampling bias of the Ly$\alpha$ sightlines,
we assign each of the densities in the sample a weight $W$ equal to
the number of redshift intervals in Table \ref{srctable} which contain
the redshift of the corresponding galaxy.  The value of $W$
therefore varies between zero ($cz>10660$~\kms) and five
($1713<cz<7901$~\kms).

When using CfA2 galaxies to trace the density field,
the survey magnitude limit causes an additional radial sampling bias.  
Even though we have corrected for the survey selection function $\psi$ in
determining $(n/\bar n)$, the CfA2 galaxies at larger redshifts 
are sampling $(n/\bar n)$ much more
sparsely because we see proportionately fewer of them per
unit volume.  We attempt to correct for this sampling bias by assigning
$(n/\bar n)$ for each galaxy an additional weighting factor of $1/\psi$.
Thus each $(n/\bar n)$ in the sample has weight $(W/\psi)$.

The weighting is not an entirely satisfactory solution to the radial bias problem,
because the unseen fraction $(1-\psi)$ of galaxies at higher redshift are 
not all located at the positions of the observed fraction $\psi$.  The
fixed smoothing length of the density estimator results in pronounced
local maxima in the density field at the locations of higher-redshift 
galaxies in the survey.  We therefore tend to overestimate the high-density
end of the $(n/\bar n)$ distribution when assuming that a particular galaxy's
$(n/\bar n)$ is representative of its $(\psi^{-1} - 1)$ unseen neighbors.

Figure \ref{allcfacdf} shows the cumulative distribution function
(CDF) of $(n/\bar n)$ for all CfA2 galaxies, with $(W/\psi)$-weighting
(dotted) as well as with {\sl W}-weighting alone (dashed).  For comparison, we also
plot the density CDF for the Ly$\alpha$ absorbers (solid).  Clearly
the Ly$\alpha$ absorbers occur far more frequently in underdense
regions.  The K-S probability that the Ly$\alpha$ CDF and the
$(W/\psi)$-weighted galaxy CDF are drawn from the same underlying
distribution is only $3\times10^{-8}$, a 5.1$\sigma$ rejection of the
null hypothesis.  The K-S probability for the {\sl W}-weighted CDF is
$9\times10^{-4}$, a 3.3$\sigma$ rejection.
\end{subsection}
\begin{subsection}{Comparison with CfA2 Galaxies in Random LOS Cylinders} 
\label{loscfa2desc}
The {\sl HST} observations of local Ly$\alpha$ absorbers sample the
population along narrow sightlines through redshift space.  This
geometric sampling is not reproduced in the volume-filling, complete
CfA2 galaxy sample (\S\ref{allcfa2desc}).  Discrete lines of sight
through the survey are less likely to sample rich galaxy clusters
because the clusters within $\sim\!100h^{-1}$~Mpc
cover only a small fraction of the sky.  We therefore define another test
population by selecting random LOS through the CfA2 survey and
extracting the galaxies within cylinders of diameter $5h^{-1}$~Mpc
(the density smoothing length) around those LOS.

For each of the
five actual AGN sightlines (Table \ref{srctable}), we assign the
corresponding redshift interval to one fifth of the randomly generated
LOS cylinders.  To obtain a suitably large test sample of 5000 galaxies, we use
$\sim\!150$ LOS.  We again correct for the selection function sampling
bias (cf.~\S\ref{allcfa2desc}) by assigning each galaxy's $(n/\bar n)$ value 
a $\psi^{-1}$ weight in the census.

Figure \ref{multicdf} shows the CDF of $(n/\bar n)$ for the
random-LOS CfA2 galaxies, $\psi^{-1}$-weighted (dotted) as well as
unit-weighted (short dash).  We again see a disparity
between the Ly$\alpha$ and CfA2 galaxy CDFs, though perhaps less
pronounced than for the full-CfA2 sample.  The sharp rise in the
unit-weighted CDF at $(n/\bar n)\approx 4$ is caused by a combination
of: (1) a preponderance of Virgo cluster galaxies, all at $(n/\bar
n)\approx 4$, appearing in the one-fifth of LOS cylinders which extend
to $z=0$ and (2) a lack of $(n/\bar n) > 4$ galaxies because of the
low probability that a $5h^{-1}$~Mpc cylinder intersects a CfA2
cluster (other than Virgo), or even a galaxy with
$cz\gtrsim8000$~\kms.  This feature is much less apparent after we
apply the selection-function weighting correction.
\placefigure{multicdf}

The K-S probability that the Ly$\alpha$ CDF and the
$\psi^{-1}$-weighted galaxy CDF are drawn from the same underlying
distribution is $3\times10^{-5}$, a 4.2$\sigma$ rejection.
The K-S probability for the unit-weighted CDF is
$2\times10^{-4}$, a 3.7$\sigma$ rejection.
\end{subsection}
\begin{subsection}{Comparison with Randomly Generated Velocities along Random LOS}
\label{maxlikedesc}
We next construct test populations of Ly$\alpha$ absorption systems
with randomly chosen redshift-space coordinates along randomly
selected LOS.  We first adopt a contstant linear density $N_{{\rm
Ly}\alpha}(cz)$ = 1/(2500~\kms) for our mock absorbers, roughly
equivalent to the observed mean incidence for the five selected {\sl HST} lines
of sight (18 absorbers over $\sim\!42000$~\kms).  Because all these
sightlines have $z_{\rm max} < 0.04$ and span a redshift range $\Delta
z \approx 0.03$, we ignore the linear density gradient 
$dN_{{\rm Ly}\alpha}(z)/dz$.
As in the previous case, we assign radial intervals to these
random LOS by selecting evenly from the five redshift windows of Table
\ref{srctable}.
\begin{subsubsection}{Single-population Model}\label{betadesc}
For a given random LOS through CfA2, we sample the smoothed galaxy density at
$1h^{-1}$~Mpc radial intervals to obtain a density-vs.-redshift profile $n(z)$.
We take this density profile,
raised to an exponent $\beta$, as the probability distribution
function for the mock-Ly$\alpha$ redshifts:
\begin{equation} \label{betaprob}
P(z)\,dz = {{n^\beta (z)\,dz}\over{\displaystyle
\int_{z_{\rm min}}^{z_{\rm max}}{n^\beta(z')\,dz'}}}.
\end{equation}
By drawing a sample of mock absorbers from a $\beta=0$ (or
$P(z) = {\rm const.}$) model, we can test the hypothesis that low column-density
Ly$\alpha$ absorbers are distributed uniformly throughout space and
completely uncorrelated with local galaxy density.  Similarly, we could
test the hypothesis that the Ly$\alpha$ systems reside in density
environments typical of CfA2 galaxies by drawing a sample of mock
absorbers from a $\beta=1$ model.  

The distribution power-law index $\beta$ may be viewed very roughly as
the combination of a cloud-creation index $\beta_{\rm c}$ and a
cloud-destruction index $\beta_{\rm d}$.  When mergers, etc., destroy
low column-density absorbers in regions of high galaxy density
($\beta_{\rm d} > 0$), we might expect an {\sl anti}correlation of
absorber location with local galaxy density ($\beta < 0$).  To
estimate $\beta$ in a merger-destruction scenario, we note that the
rate of cloud-galaxy interactions scales as the product of the local
galaxy cross section ($\propto n$) and the peculiar velocity of the
clouds through the redshift space ($\propto n^{\sim 1/2}$).  If the
clouds were produced in direct proportion to the local galaxy density
($\beta_{\rm c} = 1$), then $\beta_{\rm d} \approx 3/2$ would imply
$\beta \approx -1/2$.

Only in the case $\beta=0$ do we explicitly simulate a sample of
$\approx5000$ mock absorbers to compare with the observed set of
Ly$\alpha$ absorbers.  Figure \ref{multicdf} shows the density CDF of
the 5000 randomly located absorbers (long dash), which may be compared
with the CDF for the observed Ly$\alpha$ systems (solid).  The
agreement between the two CDFs is remarkable, particularly when
compared with the CfA galaxy CDFs in Figs.~\ref{allcfacdf} and
\ref{multicdf}.  The K-S probability that the observed Ly$\alpha$
densities share the same underlying distribution as the $\beta=0$
model is 0.20, consistent at the 1.3$\sigma$ level.  This probability
only varies by $\pm0.01$ for multiple realizations of the randomly
generated sample, indicating that a 5000-member sample is sufficiently
large for our analysis.

We check the sensitivity of this result to the adopted linear density
of randomly generated absorbers.  When we vary the linear density by
$\pm30$\%, the K-S probability varies by $\pm0.03$, larger than the
sampling variance (0.01) but still statistically insignificant.  We
also test a simulation in which the redshift interval {\sl and} the
linear density of absorbers along a random sightline are set to the
observed values for one of the five AGN sightlines.  This sampling
compensates for varying sensitivity for Ly$\alpha$ detection between
sightlines.  The analyses of the 3C 273 {\sl HST} spectra yielded
detections of lower equivalent-width absorption ($\gtrsim30$
m\AA) features than SSP96 detected for their AGN.  Not surprisingly
the 3C 273 LOS has a significantly higher linear density of absorbers
(cf.~Tables \ref{srctable} and \ref{abstable}).  The K-S probability
for this variable-linear-density model is the same as the 0.20
probability for the fixed-density model to within the sampling
variance.  The match between the density CDFs of the observed
Ly$\alpha$ clouds and the randomly chosen locations in redshift space
is robust against reasonable variation in the adopted linear density
of mock absorbers.

We next use a maximum-likelihood technique to estimate the range of
$\beta$ consistent with the observed density distribution of local
Ly$\alpha$ systems.  If equation (\ref{betaprob}) describes the actual
redshift distribution of local Ly$\alpha$ absorbers, then the
likelihood of seeing the observed redshift distribution (Table
\ref{abstable}) is the product of the individual probabilities
for each absorber at ${\bf r}_i$:
\begin{equation} \label{betalikeeq}
{\cal L}(\beta) = \prod_i{P({\bf r}_i; \beta)} = \prod_i{{{n^\beta({\bf r}_i)}\over
\displaystyle\int_{z_{i,{\min}}}^{z_{i,{\max}}}{n^\beta
(\alpha_i,\delta_i,z')\,dz'}}}
\end{equation}
Given the prior assumption that the underlying distribution of Ly$\alpha$
locations follows a $\beta$-model, then the quantity $-2 \ln[{\cal L}(\beta)
]$ should be distributed as $\chi^2$ in $\beta$.

In Figure \ref{tribeta}, the solid curve shows the model likelihood
for the range $-1.5 \leq \beta \leq 1.5$.  We obtain a maximum
likelihood value of $\beta = -0.02\pm0.23$ from the 18 nearby
Ly$\alpha$ systems.  A single-population Ly$\alpha$ forest that traces
the local large-scale structure ($\beta=1$) is inconsistent with our
result at the $4\sigma$ level.  This result may be compared with the $\sim\!4\sigma$
inconsistencies between the density CDFs of the Ly$\alpha$ absorbers and the
galaxy test populations (\S\S\ref{allcfa2desc}, \ref{loscfa2desc}), 
which should have a $\beta=1$ distribution throughout the survey.
\placefigure{tribeta}
\end{subsubsection}
\begin{subsubsection}{Dual-population Model}\label{fracdesc}
Mo and Morris (1994) speculated that the Ly$\alpha$ absorbers
may include two distinct populations: one population associated
with the gaseous extended halos of galaxies, and another population
of isolated, unclustered clouds.  In this model, we might
expect the galactic-halo fraction $f_{\rm gal}$ to have local
densities consistent with the densities around typical galaxies in the
survey ($\beta = 1$) and the unclustered fraction to have local
densities consistent with randomly chosen locations in the survey
($\beta = 0$).  Equation (\ref{betaprob}) then yields a distribution
function of dual-population absorbers of the form:
\begin{equation} \label{dualdfeq}
P(z)\,dz = {{f_{\rm gal}\,n(z)\,dz}\over{\displaystyle
\int_{z_{\rm min}}^{z_{\rm max}}{n(z')\,dz'}}} +
{{\left({1-f_{\rm gal}}\right)dz}
\over{\left({z_{\rm max}-z_{\rm min}}\right)}}.
\end{equation}
In analogy with equation (\ref{betalikeeq}) we can define a likelihood
function for the observed clouds to be drawn from a model with 
galactic fraction $f_{\rm gal}$:
\begin{equation} \label{fraclikeeq}
{\cal L}(f_{\rm gal}) = \prod_i{P({\bf r}_i; f_{\rm gal})},
\end{equation}
where $P$ is now given by equation (\ref{dualdfeq}).

In Figure \ref{trifrac}, the solid curve shows the likelihood of a dual
population model with fraction $f_{\rm gal}$ distributed like the CfA2
galaxies ($\beta=1$) and the remainder distributed uniformly
($\beta=0$).  The maximum-likelihood value for the galactic-type
fraction is $f_{\rm gal} = 0.00$, with a 1$\sigma$ upper limit of
$f_{\rm gal} = 0.24$.  This result is inconsistent at a 99\%
($2.6\sigma$) confidence level with the upper estimate of $f_{\rm gal}
\approx 0.65$ given by Lanzetta et al.~(1996).  Their lower limit of
$f_{\rm gal} \approx 0.38$ is only excluded at a confidence level of
88\% ($1.5\sigma$).
\placefigure{trifrac}
\end{subsubsection}
\end{subsection}
\end{section}

\begin{section}{Discussion} \label{discsec}
Our Ly$\alpha$ local density analysis suffers from a number of sources
of uncertainty which increase the error bars in our results of
\S\ref{results}.  Because we assign each galaxy a weight derived from
the CfA2 selection function, the density estimator is sensitive
to uncertainty in the shape of the CfA2 luminosity function.
%Furthermore, the redshift survey of Zwicky galaxies is
%incomplete beyond the boundaries of the CfA2 survey, leading to a
%systematic depression of the estimated density within roughly one
%smoothing length ($5h^{-1}$~Mpc) of the survey edges.  
%Because we construct the smoothed density field from a point
%distribution of galaxies, 
The density estimator, constructed from a point distribution of
galaxies, is subject to increasing shot noise at larger redshift.  The
shot noise dominates at $cz\gtrsim 10000$~\kms, where the mean
distance between survey galaxies rapidly exceeds our $5h^{-1}$~Mpc
smoothing length.  For this reason, we may worry that the parameter
constraints are also sensitive to the adopted smoothing length.  Here
we quantify some of these sources of uncertainty and estimate the
improvements in Ly$\alpha$ distribution constraints which would result
from a larger sample of nearby absorbers.

\begin{subsection}{Uncertainty in the CfA2 Luminosity Function} \label{lfdisc}
The uncertainty in the CfA2 luminosity function 
(MHG94) introduces an uncertainty in our selection
function weighting (eq. \ref{selfneq}).  Our results for
$\beta$ and $f_{\rm gal}$ are insensitive to the overall LF
normalization, $\phi_*$; it cancels out of equations
(\ref{betaprob}) and (\ref{dualdfeq}).  We may therefore restrict our
error analysis to uncertainty in the LF shape parameters,
$\alpha$ and $M_*$.  To incorporate this LF uncertainty into our error
bars on $\beta$ and $f_{\rm gal}$, we Monte Carlo resample the 
densities in equations
(\ref{betaprob}) and (\ref{dualdfeq}), varying the LF
parameters according to their ranges from MHG94: $M_* = -18.8\pm0.3$,
and $\alpha = -1.0\pm0.2$.  The 1$\sigma$ bounds on the
absorber distribution parameters increase only marginally from
\S\ref{results}:
\begin{equation} \label{varlfeq}
\beta : -0.02\pm0.23 \to -0.02\pm0.24;\quad f_{\rm gal} : 0.00^{+0.24}_{-0}
\to 0.00^{+0.25}_{-0}.
\end{equation}

MHG94 show that the respective LFs of CfA2 North and CfA2 South differ
significantly ($>2\sigma$) with the LF of the Combined (CfA2
North+South) sample.  This North/South LF difference is difficult to
interpret; it may be caused by large-scale systematic errors in the
Zwicky photometry or by real variations in the underlying LF across
the survey volume.  We investigate its effect on our density
distribution analysis by replacing the selection function derived from
the Combined sample ($\phi_*=0.04\,({\rm Mpc}/h)^{-3}$,\ $M_* =
-18.8$, $\alpha=-1.0$) with a hybrid selection function using the LFs
from CfA2 North ($\phi_*=0.05\,({\rm Mpc}/h)^{-3}$, $M_* = -18.67$,
$\alpha=-1.03$) and CfA2 South ($\phi_*=0.02\,({\rm Mpc}/h)^{-3}$,
$M_* = -18.93$, $\alpha=-0.89$) as appropriate.  The hybrid selection
function results in insignificant changes to the model parameters:
\begin{equation} \label{hybridlfeq}
\beta : -0.02\pm0.23 \to -0.05\pm0.26;\quad f_{\rm gal} : 0.00^{+0.24}_{-0}
\to 0.00^{+0.25}_{-0}.
\end{equation}
Based upon Equations (\ref{varlfeq}) and (\ref{hybridlfeq}), we conclude that
our density neighborhood technique is highly
robust against uncertainty in the survey LF.

\end{subsection}
\begin{subsection}{Choice of Smoothing Length} \label{smoothdisc}
To determine the sensitivity of our results to the particular galaxy
smoothing length, we repeat the analysis of \S\ref{results} with
smoothing kernels of 2.5$h^{-1}$~Mpc and 10$h^{-1}$~Mpc.  These
values are near the functional limits on kernel size: broader kernels
wash out the large-scale structure we are hoping to probe, and
narrower kernels are subject to excessive shot noise.

Figures \ref{tribeta} and \ref{trifrac}, respectively, show the
$\beta$ and $f_{\rm gal}$ likelihood curves for kernel smoothing
lengths of 2.5$h^{-1}$~Mpc (dashed), 5$h^{-1}$~Mpc (solid), and
10$h^{-1}$~Mpc (dotted).  As might be expected, the low-contrast
10$h^{-1}$~Mpc smoothing length results in poorer parameter
constraints; the high contrast of the 2.5$h^{-1}$~Mpc smoothing gives
tighter constraints.  These curves do not include shot noise
corrections (\S\ref{shotsec}), which have a larger broadening effect
at the smaller smoothing length.  We give the 1$\sigma$ confidence
intervals on $\beta$ and $f_{\rm gal}$ for the three smoothing lengths
in Table \ref{trisigtab}.  For the range of smoothing length
2.5--10$h^{-1}$~Mpc, the maximum likelihood values for $\beta$ and
$f_{\rm gal}$ remain consistent with 0 (randomly distributed
absorbers) at the $1.2\sigma$ level or less.  Likewise, this range
of smoothing lengths remains inconsistent with $\beta=1$ (absorbers
tracing structure) at the $3.8\sigma$ level or more.  Thus our
results are largely insensitive to the particular choice of smoothing
kernel scale over a reasonable range.
\placetable{trisigtab}
\end{subsection}
\begin{subsection}{Shot Noise in the Density Estimator} \label{shotsec}
As a first estimate of the $(n/\bar n)$ shot noise, we construct a
mock redshift survey drawn from the CfA2 LF.  We place the mock survey
galaxies at random locations in redshift space, with the underlying
density $(n/\bar n) \equiv 1$ throughout.  We then remove galaxies
fainter than the CfA2 magnitude limit and use the remaining sample to
estimate the density contrast throughout the survey volume.  By
construction, the $(n/\bar n)$ shot noise throughout this homogeneous
survey should only be a function of redshift.  Figure \ref{mockvar}
shows the variance of the density estimate, binned in $2.5h^{-1}$~Mpc
radial intervals out to $150h^{-1}$~Mpc.  Shot noise rapidly
dominates the density estimate for $cz \gtrsim 12000$~\kms, where the
galaxies on average are separated by more than twice the $5h^{-1}$~Mpc
smoothing length.  Thus we restrict our analysis to Ly$\alpha$
absorption systems with $cz \lesssim 10500$~\kms.  The
systematic decline in the median estimated density at random locations
with $cz\gtrsim10000$~\kms\ has virtually no influence on our test
population of random locations (\S\ref{betadesc}), drawn from the
$cz_{\rm max}\lesssim10500$~\kms\ radial intervals of Table \ref{srctable}.
\placefigure{mockvar}

Unlike our mock survey, CfA2 is not homogeneous.  The true shot noise
will be relatively larger in regions of low galaxy number density, and
vice versa.  We reproduce this structure-dependent shot noise by
bootstrap resampling the entire redshift dataset.  We recompute the
likelihood $\beta$ and $f_{\rm gal}$ curves (eqs. \ref{betaprob} and
\ref{dualdfeq}) for 1024 bootstrap realizations of the redshift
dataset.  The $1\sigma$ bounds expand to $-0.26 <
\beta < 0.22$, and $0.00 < f_{\rm gal} < 0.25$, as compared with the
ranges in \S\ref{maxlikedesc}.  This minor increase grows large if we
attempt to include the next two nearest Ly$\alpha$ systems, which are
in the 3C 273 foreground at $cz\approx 15000$~\kms.  The large
uncertainty ($\pm1$~dex) in the density estimate at 15000~\kms\
overwhelms any benefit to be gained from the larger sample size.
In \S\ref{deep3c}, we discuss the observational requirements to
include the 3C 273 absorbers at $z\gtrsim0.05$ in our analysis.
\end{subsection}
\begin{subsection}{Improving the Parameter Constraints with a Larger Sample}
\label{deep3c}
With 18 Ly$\alpha$ absorbers, the distribution of surrounding galaxy
density contrasts places interesting constraints upon the nature of
the low-redshift Ly$\alpha$ forest.  A larger sample of absorber local
densities should further shrink the error bars on the parameter
constraints.  Furthermore, local Ly$\alpha$ absorber searches along
additional LOS would improve the likelihood that the cumulative
pathlength probes a representative distribution of the galaxy number
density within the survey.

We quantify the improved constraints with a Monte Carlo simulation
involving 50 mock absorbers placed along random LOS with the same mean
frequency as our current sample.  We assign redshifts to the mock
absorbers according to an {\it a priori} $\beta$- or $f_{\rm
gal}$-model.  The density contrast analysis recovers the seed values
$\beta$ or $f_{\rm gal}$ with 1$\sigma$ error bars of only
$\sim\!0.10$.  Were the nearby Ly$\alpha$ absorbers truly distributed
randomly, for example, a distribution of density contrasts for 50
absorbers should conclusively rule out a model with $\beta=1$ at the
$9\sigma$ level.  Assuming a local absorber frequency of
$\sim\!1/2500$~\kms\ detectable with {\sl HST}/STIS, a sample of
50 absorber densities would require roughly a dozen AGN
lines of sight within surveyed regions.

The current limits to the sample size of local Ly$\alpha$ absorber
densities are: (1) the small number of AGN sightlines with identified
low-redshift absorbers and (2) the limited depth of the complete
redshift surveys for density estimation.  As an example of the latter,
there are ten more 3C 273 absorbers, with $0.05
\lesssim z \lesssim 0.16$, whose local densities cannot be estimated
from a $m_{\rm Zw}\leq15.5$ redshift survey.  Including all the
additional 3C 273 absorbers would increase the sample size to 28 and
almost double the cumulative pathlength currently probed by the
various LOS in this study.  Below we make a rough estimate of the
observational requirements for studying the large-scale density
environment around these more distant absorbers in a manner consistent
with our current sample.  Our technique differs substantially from the
approach suggested by Sarajedini, Green, \& Jannuzi (1996) for the
even more distant ($0.2<z<0.4$) absorbers detected with {\sl HST}/FOS.

To determine the $(n/\bar n)$ profile along the entire 3C 273 LOS
without significant edge dampening or shot noise, we need to
survey galaxies brighter than $\sim\!M_*$ out to $\sim\!2$ smoothing
lengths, or $10h^{-1}$~Mpc, from the LOS.  At $z=0.05$ this
corresponds to $B\sim16.5$ over $\sim\!50$ square degrees,
decreasing to $B\sim19.5$ over $\sim\!5.6$ square degrees at $z=0.16$.
For comparison, Morris et al.\ (1993) have already surveyed
$\sim\!3.5$ square degrees around 3C 273 to $B\sim19$, and Grogin et
al.\ (1998) have recently completed a $2^{\rm h}\times12\arcdeg$
redshift survey around 3C 273 to the Zwicky catalog limit ($m_{\rm Zw}
= 15.7$).
\end{subsection}
\end{section}

\begin{section}{Conclusions}\label{concsec}
We develop a new diagnostic for exploring the relationship between
Ly$\alpha$ absorption systems and the galaxy distribution: we compare
the CDF of the smoothed galaxy number density at the absorber
locations with the density CDF for test populations within the
redshift survey.  The underlying assumption is that if Ly$\alpha$
absorbers sample the large-scale structure revealed by redshift
surveys, the distribution of smoothed galaxy number densities around
absorber positions in redshift space should be consistent with
densities around galaxies in the redshift survey.

The procedure for evaluating the smoothed density field is
straightforward, requiring only the galaxy positions and
redshifts, the survey selection function, and a choice of smoothing
kernel.  Moreover, this technique may be easily generalized as long as
the redshift survey dimensions are large compared to the smoothing
kernel.  For sensitivity to large-scale structure without excessive
blurring of the survey density contrasts, we adopt a Gaussian
smoothing kernel of width $5h^{-1}$~Mpc, which is approximately the
galaxy-galaxy correlation length.

The density CDF of 18 Ly$\alpha$ absorbers within regions surveyed to
$m_{Zw}=15.5$ is inconsistent with the comparable CDF for CfA2
galaxies.  The K-S probability that the two CDFs are drawn from the
same underlying distribution never exceeds $2\times10^{-4}$ for a
variety of CfA2 sampling geometries and weighting schemes
(\S\S\ref{allcfa2desc}, \ref{loscfa2desc}).  The low probability
results from a much larger fraction of Ly$\alpha$ absorbers than
galaxies in underdense regions of the redshift survey.  By contrast, a
CDF of densities at randomly selected locations in the survey
represents the same underlying density distribution as the Ly$\alpha$
absorbers with a K-S probability of 20\%.

We explore the likelihood that the nearby Ly$\alpha$ absorbers are:
(1) a single population distributed as the galaxy density raised to
some power-law $\beta$, or (2) a dual population of absorbers, with a
fraction $f_{\rm gal}$ distributed like the galaxies ($\beta=1$) and
the remainder distributed uniformly with respect to the galaxies
($\beta=0$).  For the 18 absorbers in this study, the $\pm1\sigma$
likelihood interval of $\beta$ is $-0.02\pm0.26$.  The error bars
include the uncertainty in the density estimator from the survey
sampling variance and the uncertain survey LF, both small
contributions.  The maximum likelihood value of $f_{\rm gal}$ is 0.0,
with a $1\sigma$ upper limit of 0.26.  This value is consistent with
the $f_{\rm gal}\sim0.20$ found by Mo \& Morris (1994) for the 3C 273
absorbers alone, but at odds with the $f_{\rm gal}=0.65\pm0.18$ found
by Lanzetta et al.~(1995) for a sample of {\sl HST}/FOS absorbers at
higher column density.

We conclude that the low redshift, low column density Ly$\alpha$
clouds in this sample are {\sl not} tracing the nearby large-scale
structure marked by typical luminous galaxies.  The absorbers appear
in a range of density environments similar to those around randomly
chosen locations throughout the survey.  Thus the $N_{H{\sc
i}}\lesssim10^{14}$~cm$^{-2}$ clouds at $z<0.035$ are apparently
similar in their spatial distribution to the unclustered, low column
density clouds seen at high $z$ (\cite{sarg80}).  The sharp decline in
absorber metallicity at $N_{H{\sc i}}\lesssim10^{14}$ recently
reported by Lu et al.~(1998) suggests that the absorbers found nearby
with {\sl HST}/GHRS may be a distinct population from the {\sl
HST}/FOS absorber sample, which Lanzetta et al.~(1995) argue is
associated ($f_{\rm gal}\sim0.65$) with $\sim160h^{-1}$~kpc galactic
halos.

Although our sample is large enough to reject $f_{\rm gal} = 1$, it
does not exclude the conservative lower limit, $f_{\rm
gal}=0.35\pm0.10$, of Lanzetta et al.~(1995).  In order to reduce the
$\beta$ and $f_{\rm gal}$ error bars to $\pm0.10$, the precision
necessary to exclude the Lanzetta et al.\@ lower limit, Monte Carlo
simulations indicate that a sample of $\sim\!50$ nearby absorbers is
required.  The additional $\sim10$ sightlines needed for a 50-absorber
sample would also increase the likelihood that the cumulative
pathlength intersects the full range of density environments in the
nearby universe.  With the progress of wide-field imaging and
multi-object spectroscopy, redshift surveys around the {\sl HST}
Ly$\alpha$ absorber sightlines using 4m-class telescopes should soon
enable the density environment analysis presented here to encompass
many more Ly$\alpha$ absorbers out to $z\lesssim 0.4$.  Investigation
of absorber density CDFs for subsamples grouped by column density
might then reveal a contrast in absorber spatial distribution akin to
the metallicity falloff seen by Lu et al.~(1998).
\acknowledgments
We thank Susan Tokarz, Emilio Falco, and Michael Kurtz for assistance
with the CfA2 survey database, and Jan Kleyna for useful discussions.
This research is supported in part by the Smithsonian Institution.
\end{section}
\clearpage
\begin{table}
\caption[]{ \label{srctable}
Targets of Local Ly$\alpha$ Absorption Searches }
\smallskip
\begin{tabular}{lcccr@{--}l}\tableline \tableline
Target & R.A. & Dec. & Redshift\tablenotemark{a} & 
\multicolumn{2}{c}{Ly$\alpha$ Sensitivity\tablenotemark{a}} \\
& (B1950) & (B1950) & (\kms) & \multicolumn{2}{c}{(\kms)}\\ \tableline
Mrk 335 & $00^{\rm h}03^{\rm m}45\fs2$&$+19\arcdeg55'29''$ 
& 7901 & 1713 & 7901 \\
I Zw 1  & $00^{\rm h}50^{\rm m}57\fs8$&$+12\arcdeg25'20''$ 
& 18490 & 1660 & 10660 \\
Mrk 421 & $11^{\rm h}01^{\rm m}40\fs6$&$+38\arcdeg28'43''$ 
& 9001 & 1501 & 9001 \\
3C 273  & $12^{\rm h}26^{\rm m}33\fs2$&$+02\arcdeg19'43''$ 
& 47347 & 0 & 10500\tablenotemark{b} \\
Mrk 501 & $16^{\rm h}52^{\rm m}11\fs7$&$+39\arcdeg50'25''$ 
& 10301 & 1709 & 10301 \\ 
\tableline
\end{tabular}
\tablenotetext{a}{Galactocentric velocities.}
\tablenotetext{b}{Absorbers detected to $z\sim0.12$, but
CfA2 too sparse for meaningful density estimation.}
\end{table}

\begin{table}
\caption[]{ \label{abstable}
Local Ly$\alpha$ Absorption Velocities and Equivalent Widths}
\smallskip
\begin{tabular}{lcc@{\qquad}lcc} \tableline\tableline
Background & $cz_{\rm abs}$ & $W_\lambda$ & Background & $cz_{\rm abs}$ & $W_\lambda$ \\
Source & (\kms) & (m\AA) & Source & (\kms) & (m\AA) \\ \tableline
Mrk 335 & 2183 & 170 & 		3C 273 	&2060 & 240\\
& 2503 & 73 &				&6413 &27\\
& 4483 & 26 &				&7737 &32\\
& 6493 & 140 &				&8704 &120\\
I Zw 1 & 1777 & 120 &			&9750 &74\\
& 3021 & 84 &			Mrk 501 &4869 & 154\\	
& 5290 & 84 &				&6209 & 36\\	
Mrk 421 & 3047 & 92 &			&7739 & 48\\
3C 273 & 890 & 371&			& & \\
&1460 &414&				& & \\
\tableline
\end{tabular}
\tablecomments{Velocities are galactocentric. 3C 273 data from
Morris et al.~(1991), Bahcall et al.~(1991), and Weymann
et al.~(1995); remaining data from Shull, Stocke, \& Penton
(1996).}
\end{table}

\begin{table}
\caption[]{ \label{trisigtab}
Results for Various Smoothing Kernel Scales}
\smallskip
\begin{tabular}{cccc} \tableline\tableline
Param. & \multicolumn{3}{c}{Kernel Smoothing Length} \\
& 2.5 Mpc & 5 Mpc & 10 Mpc \\ \tableline
$\beta$ & $0.09\pm0.13$ & $-0.02\pm0.23$ & $-0.45\pm0.38$ \\
$f_{\rm gal}$ & $0.06^{+0.20}_{-0.06}$ & $0.00^{+0.24}_{-0.00}$ & 
$0.00^{+0.24}_{-0.00}$ \\
\tableline
\end{tabular}
\tablecomments{1$\sigma$ (68.3\%) confidence intervals, without
corrections for shot noise or LF uncertainty.}
\end{table}

\clearpage
\begin{center} {\large Figure Captions} \end{center}
\medskip
\figcaption[3c273clip.ps]{ 
Redshift survey slice and density profile along 3C 273 line of sight.  See
\S\ref{sliceplots} for description.
\label{3cwedge}}

\figcaption[335clip.ps]{ 
Redshift survey slice and density profile along Mrk 335 line of sight.  See
\S\ref{sliceplots} for description.
\label{335wedge}}

\figcaption[421clip.ps]{ 
Redshift survey slice and density profile along Mrk 421 line of sight.  See
\S\ref{sliceplots} for description.
\label{421wedge}}

\figcaption[501clip.ps]{ 
Redshift survey slice and density profile along Mrk 501 line of sight.  See
\S\ref{sliceplots} for description.
\label{501wedge}}

\figcaption[izw1clip.ps]{ 
Redshift survey slice and density profile along I Zw 1 line of sight.  See
\S\ref{sliceplots} for description.
\label{izw1wedge}}

\figcaption[cfacdf.ps]{
Cumulative distribution function of galaxy density contrast around 18
nearby Ly$\alpha$ absorbers (solid), and around all CfA2 galaxies.
The CfA2 sample's CDF is shown with (dotted curve) and without (dashed
curve) selection-function weighting (cf.~\S\ref{allcfa2desc}).
\label{allcfacdf}}

\figcaption[multicdf.ps]{
Cumulative distribution functions of galaxy density contrast for
various samples probing sightlines through the CfA survey, including: 
18 nearby Ly$\alpha$ absorber locations from five AGN sightlines
(solid); CfA2 galaxies within $2.5h^{-1}$~Mpc of random sightlines,
with (dotted) and without (short dash) selection-function weighting
(cf.~\S\ref{loscfa2desc}); and random redshifts from random sightlines,
with the same mean redshift spacing as the Ly$\alpha$ sample
(long dash).
\label{multicdf}}

\figcaption[tribeta.ps]{
Likelihood curves for models with Ly$\alpha$ absorbers distributed as
a power law $\beta$ of the galaxy density contrast
(cf.~\S\ref{betadesc}).  Results shown for galaxy smoothing lengths of
$2.5h^{-1}$~Mpc (dotted), $5h^{-1}$~Mpc (solid), and $10h^{-1}$~Mpc
(dashed).
\label{tribeta}}

\figcaption[trifrac.ps]{
Likelihood curves for models with Ly$\alpha$ absorbers distributed in
two populations, with a fraction $f_{\rm gal}$ distributed as the CfA2
galaxies and the remainder distributed randomly
(cf.~\S\ref{fracdesc}).  Results shown for galaxy smoothing lengths of
$2.5h^{-1}$~Mpc (dotted), $5h^{-1}$~Mpc (solid), and $10h^{-1}$~Mpc
(dashed).
\label{trifrac}}

\figcaption[mockvar.ps]{
Variance in the $5h^{-1}$~Mpc density contrast, $(n/\bar n)$, within a
simulated CfA2 survey of uniform density throughout (dotted line).
The median value (crosses) and $\pm1\sigma$ bounds are shown for the
density contrast sampled at a large ($N>1000$) number of random
locations within successive $2.5h^{-1}$~Mpc radial shells at distance $D$.  
Because of the growing systematic and statistical errors in the density 
estimator at $cz\gtrsim12000$~\kms, we limit our analysis to Ly$\alpha$ absorbers
with $cz<10500$~\kms.
\label{mockvar}}
\clearpage
\pagestyle{empty}
\begin{figure}[bp]
\begin{center} {\large FIGURE 1}\end{center}
\plotone{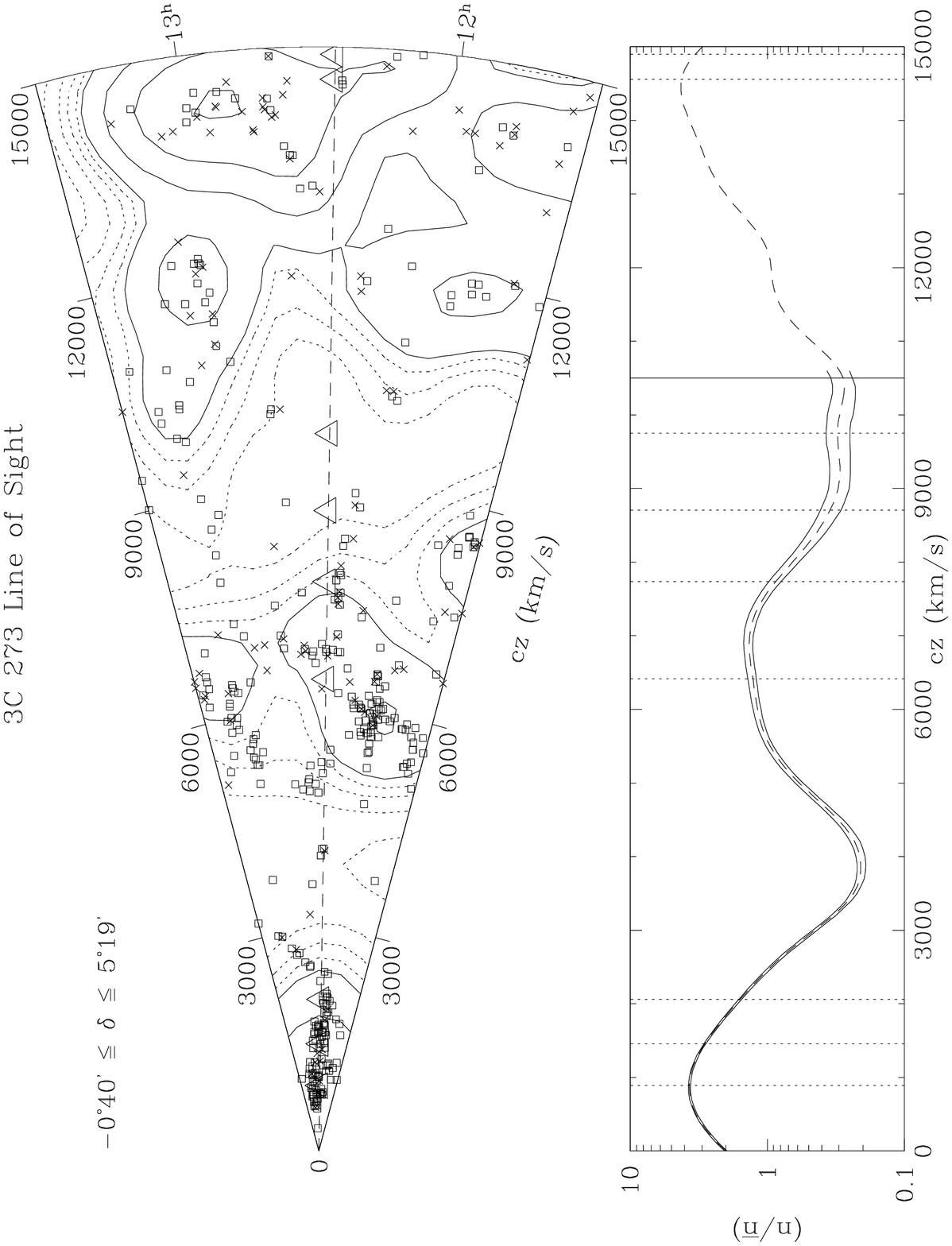}
\end{figure}
\begin{figure}[bp]
\begin{center} {\large FIGURE 2}\end{center}
\plotone{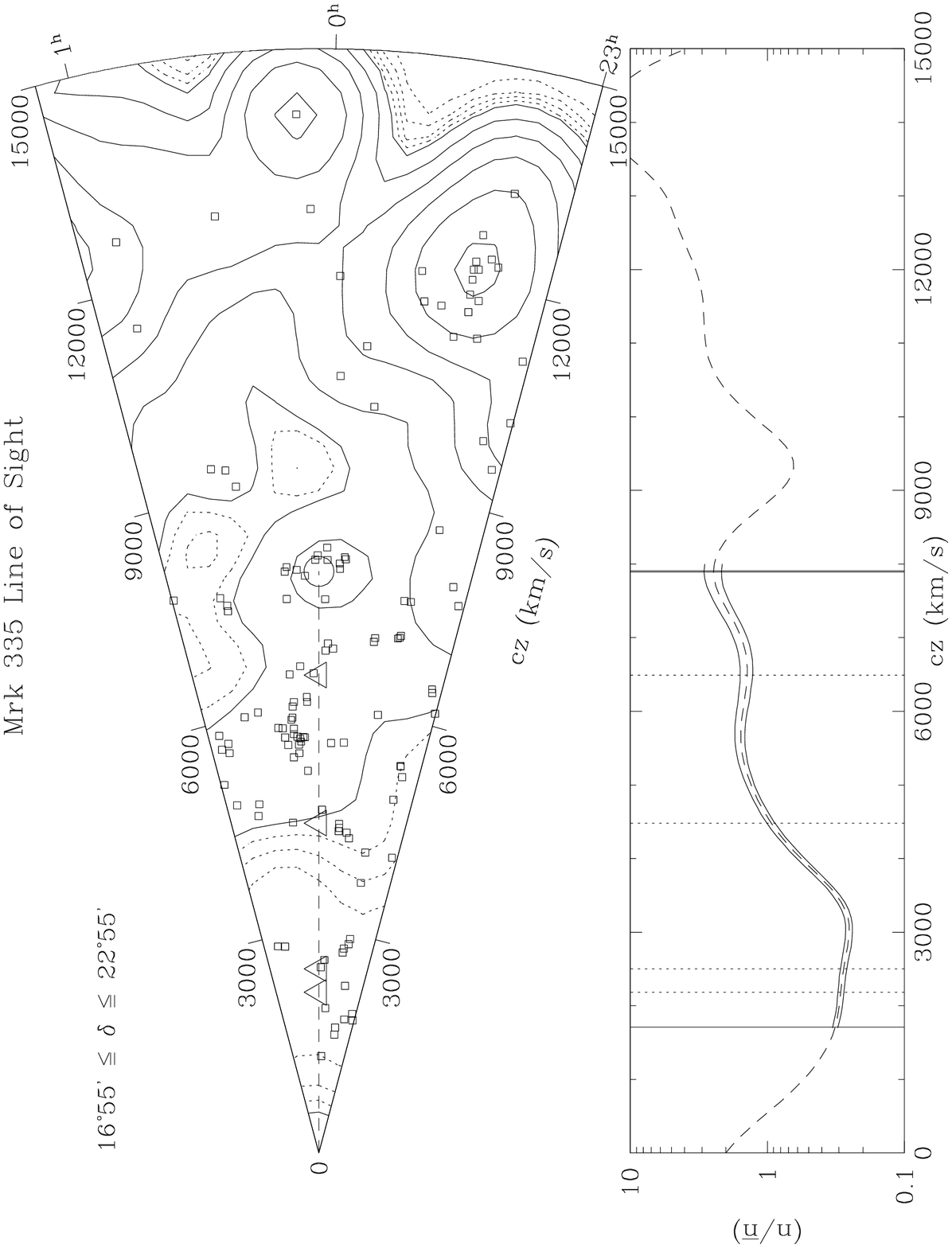}
\end{figure}
\begin{figure}[bp]
\begin{center} {\large FIGURE 3}\end{center}
\plotone{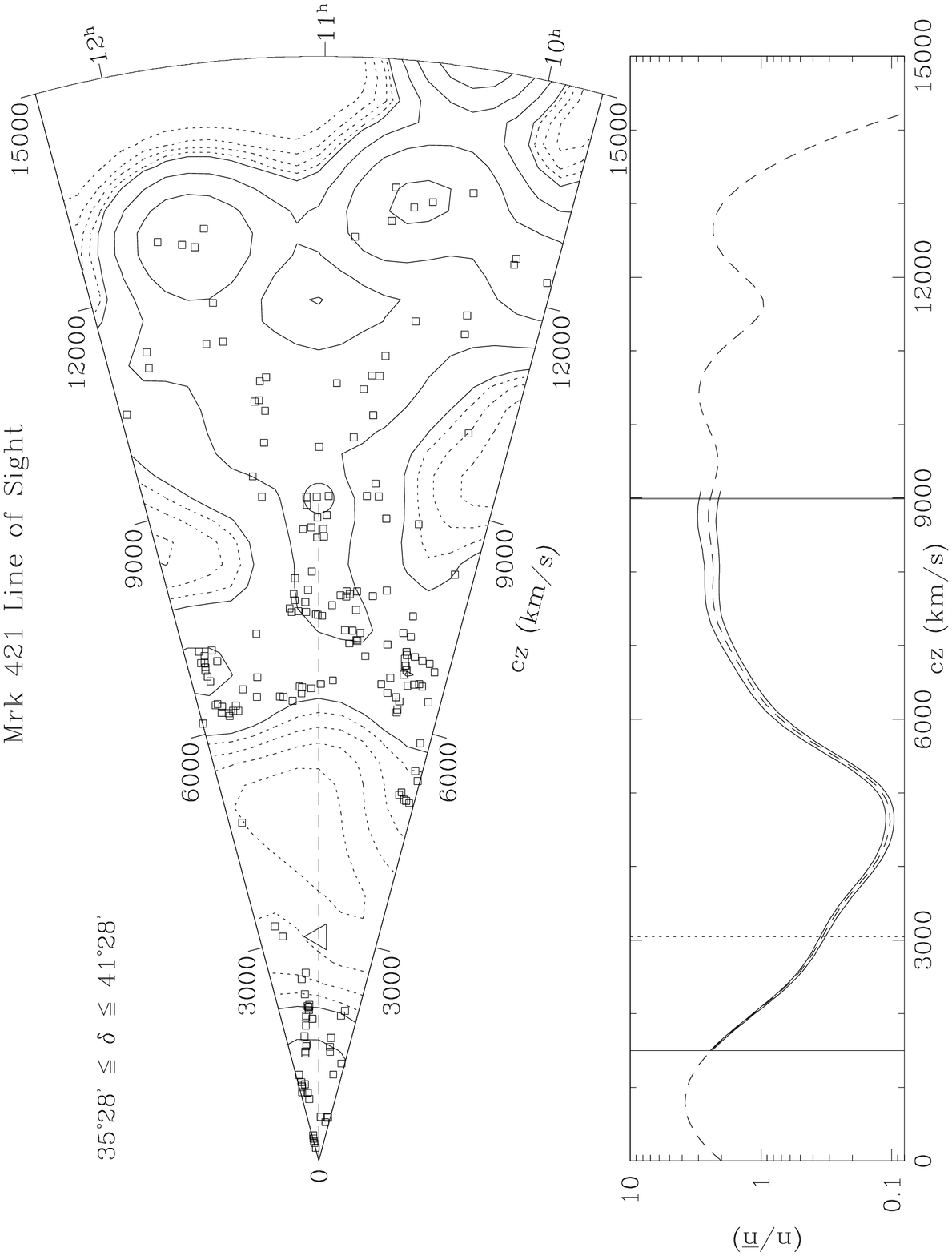}
\end{figure}
\begin{figure}[bp]
\begin{center} {\large FIGURE 4}\end{center}
\plotone{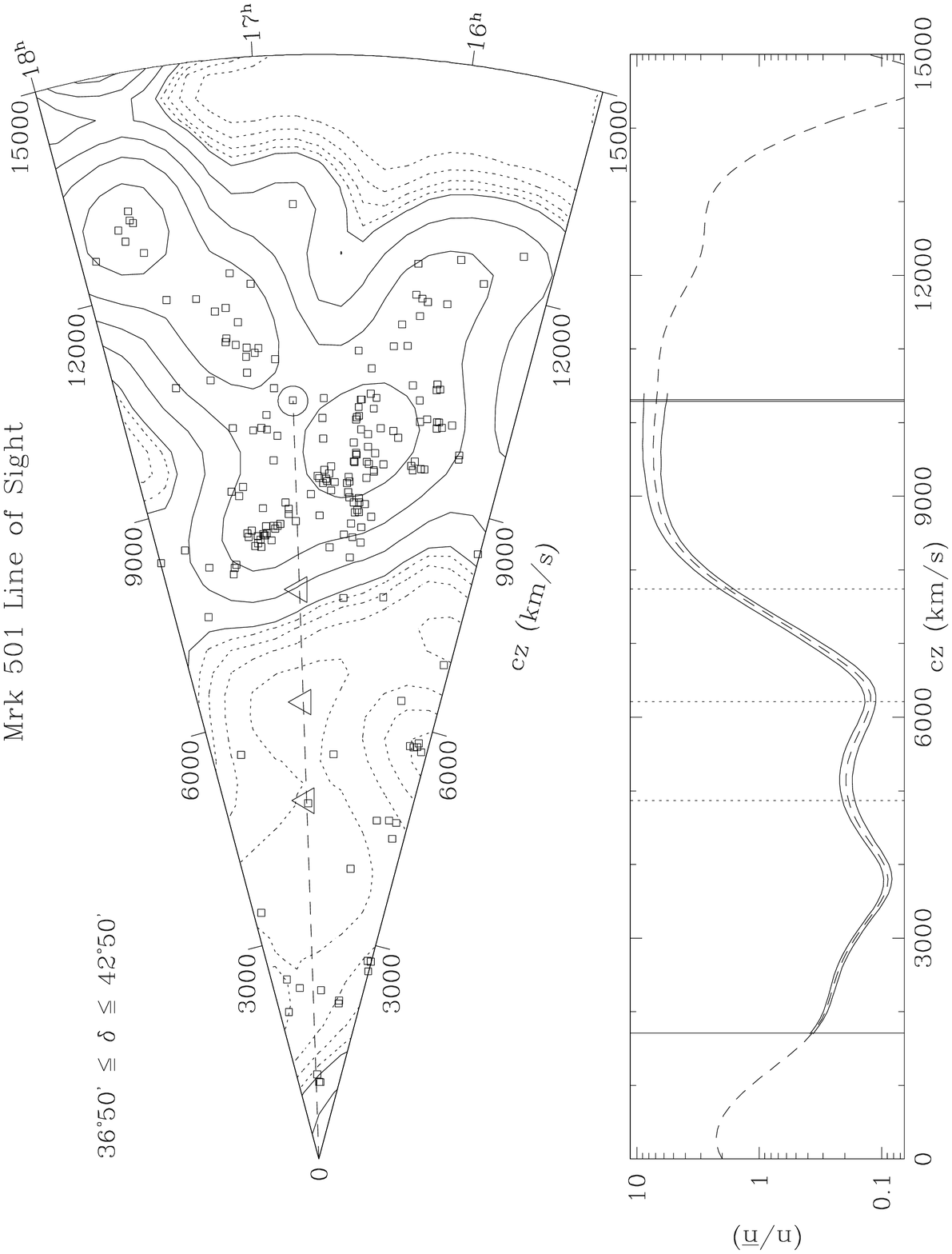}
\end{figure}
\begin{figure}[bp]
\begin{center} {\large FIGURE 5}\end{center}
\plotone{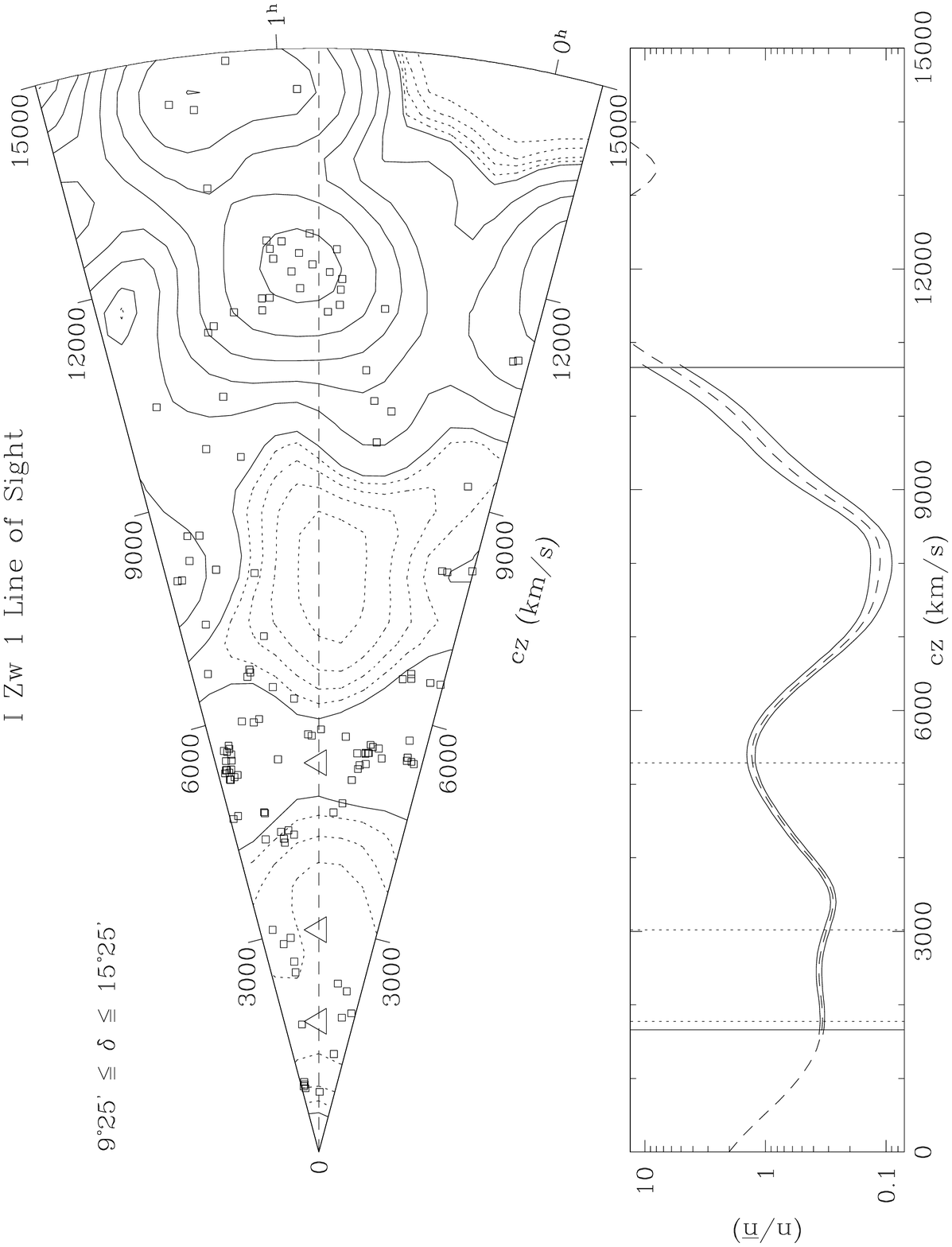}
\end{figure}
\begin{figure}[bp]
\begin{center} {\large FIGURE 6}\end{center}
\plotone{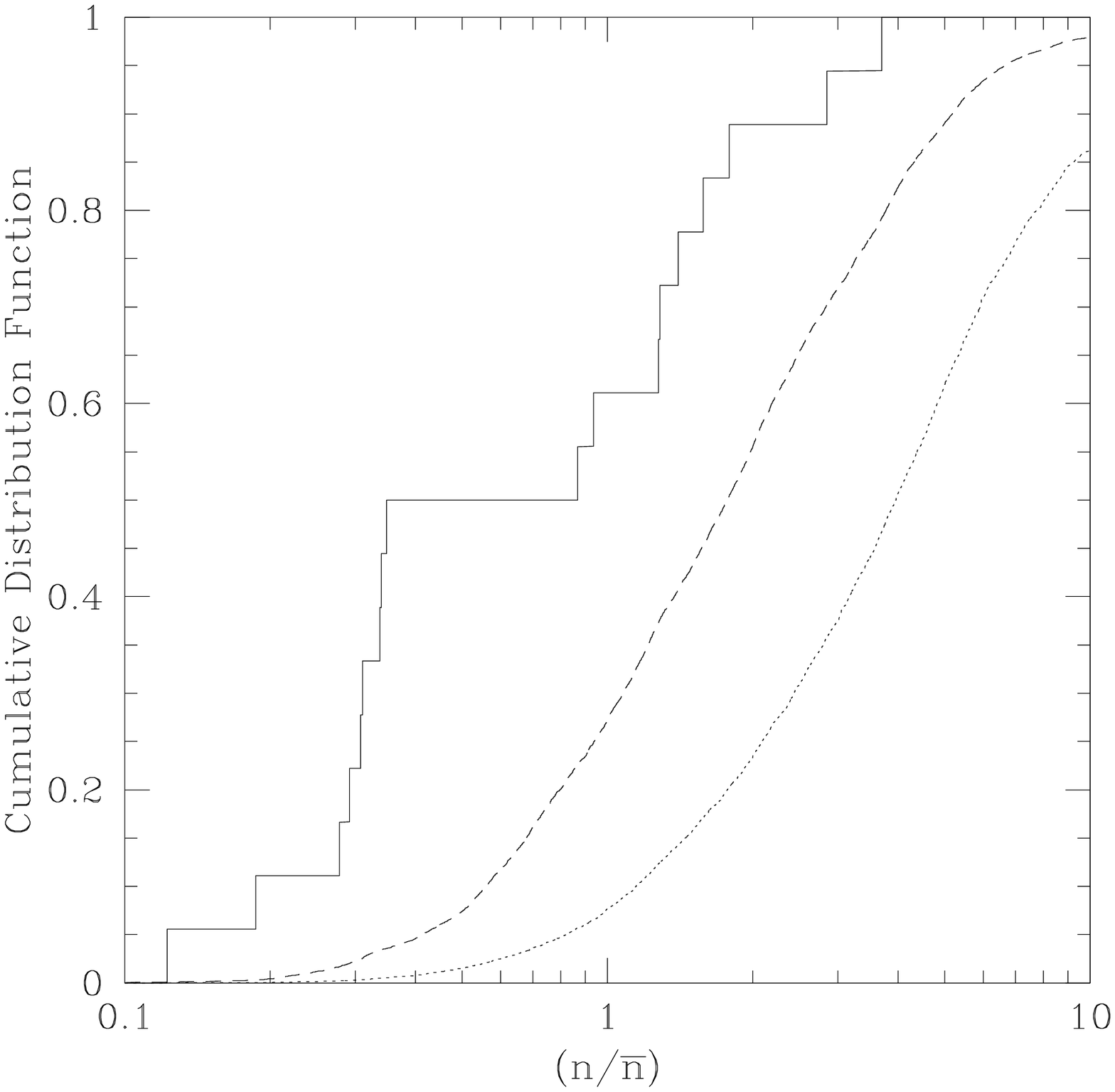}
\end{figure}
\begin{figure}[bp]
\begin{center} {\large FIGURE 7}\end{center}
\plotone{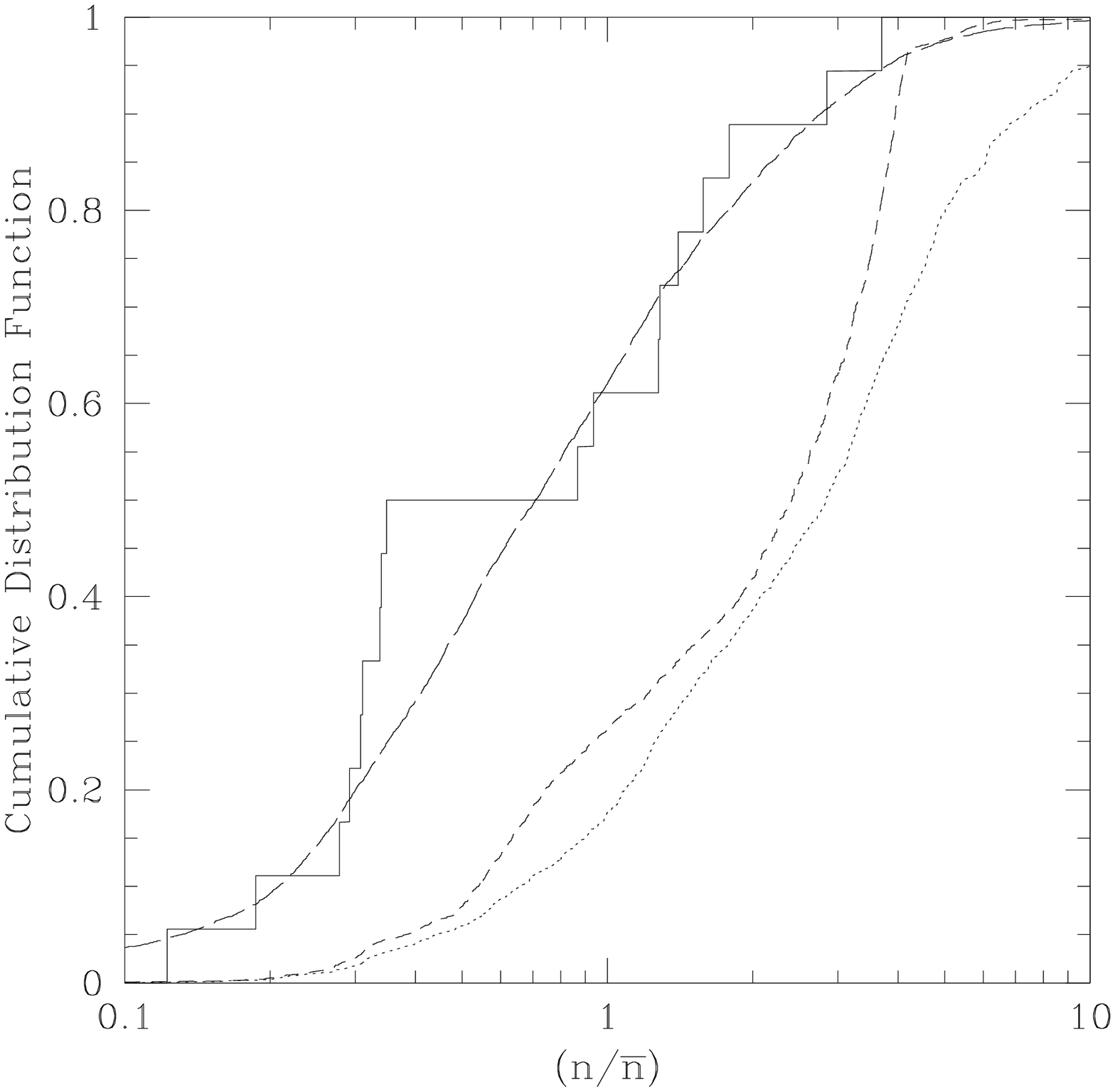}
\end{figure}
\begin{figure}[bp]
\begin{center} {\large FIGURE 8}\end{center}
\plotone{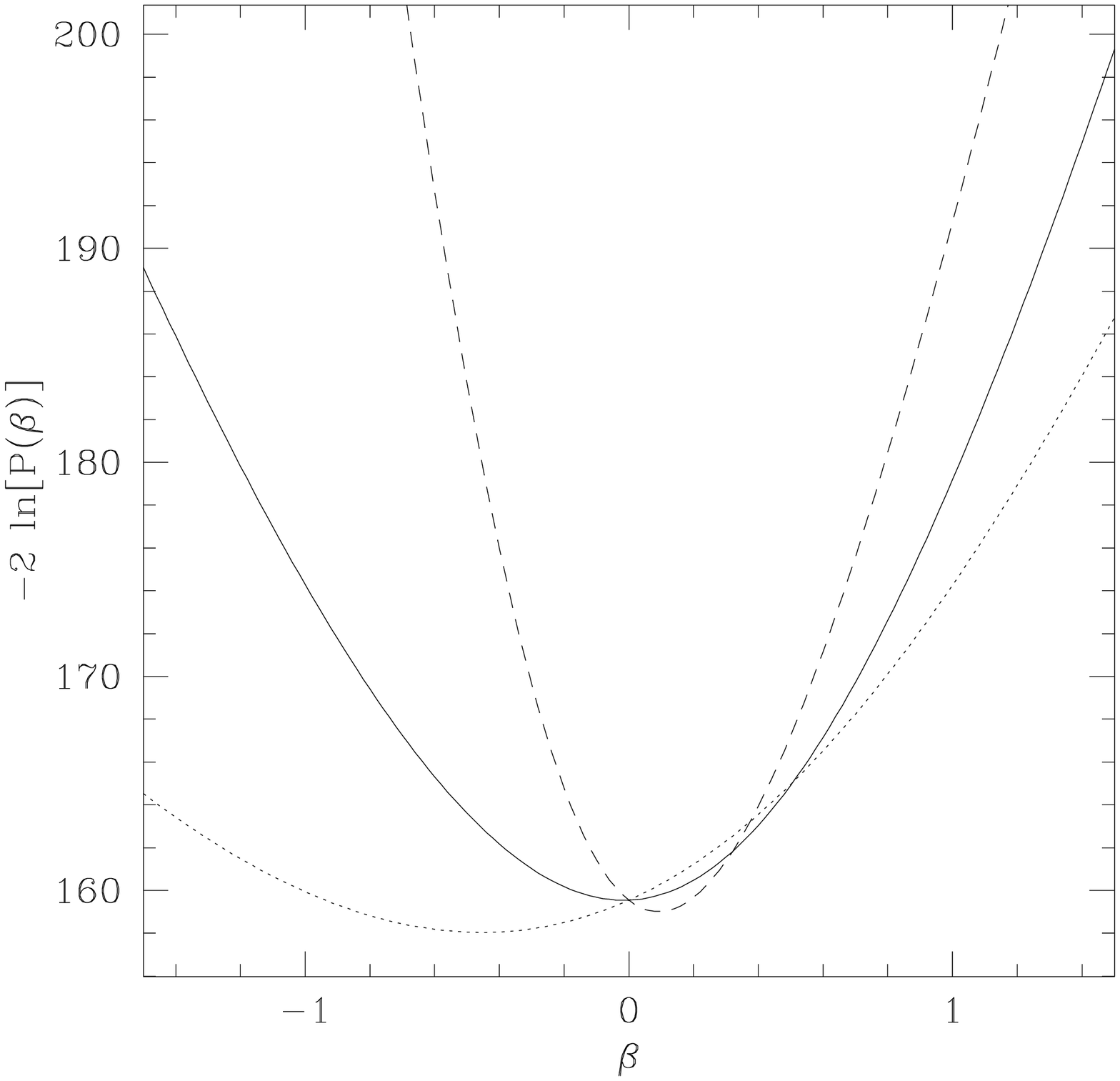}
\end{figure}
\begin{figure}[bp]
\begin{center} {\large FIGURE 9}\end{center}
\plotone{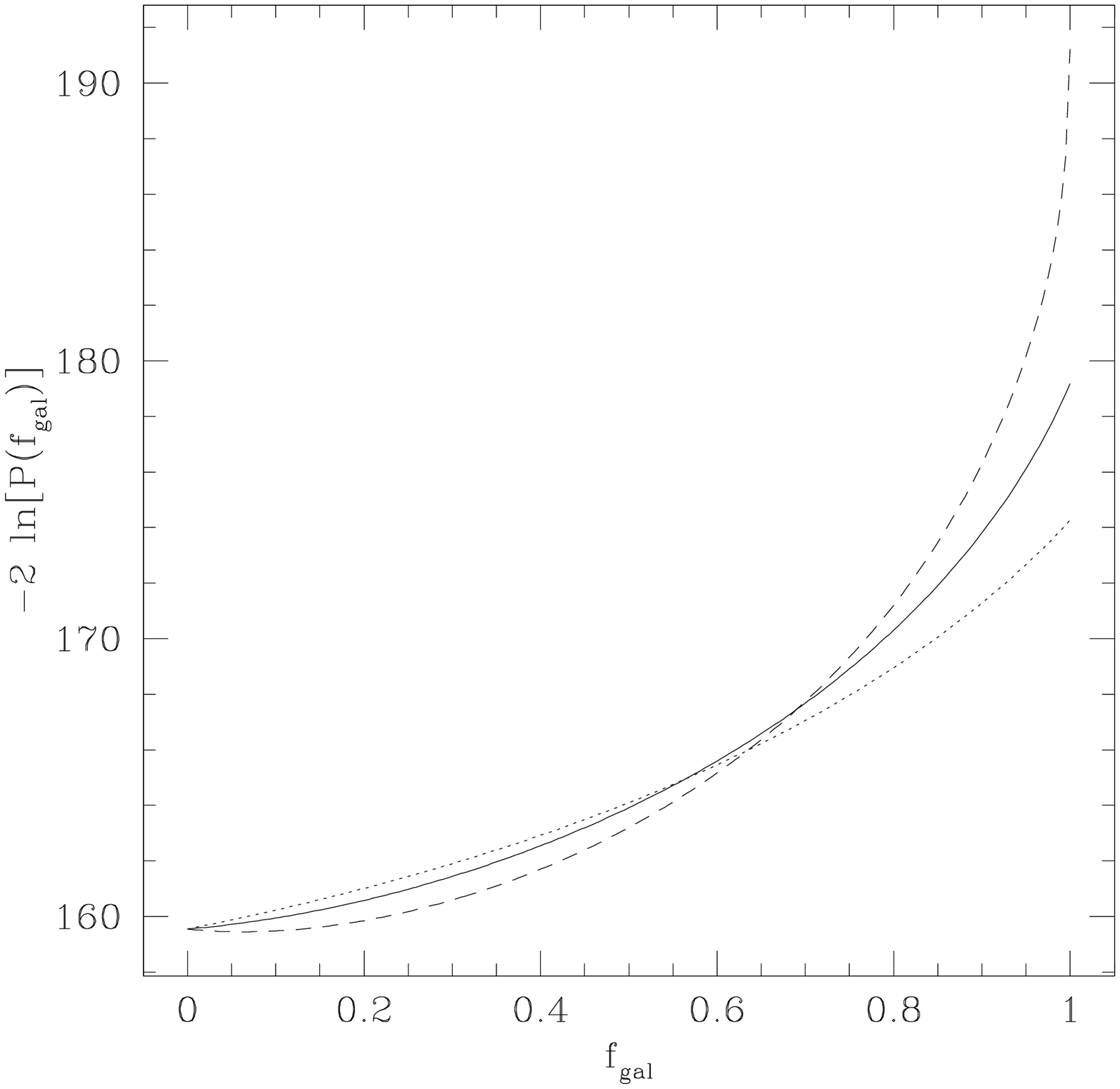}
\end{figure}
\begin{figure}[bp]
\begin{center} {\large FIGURE 10}\end{center}
\plotone{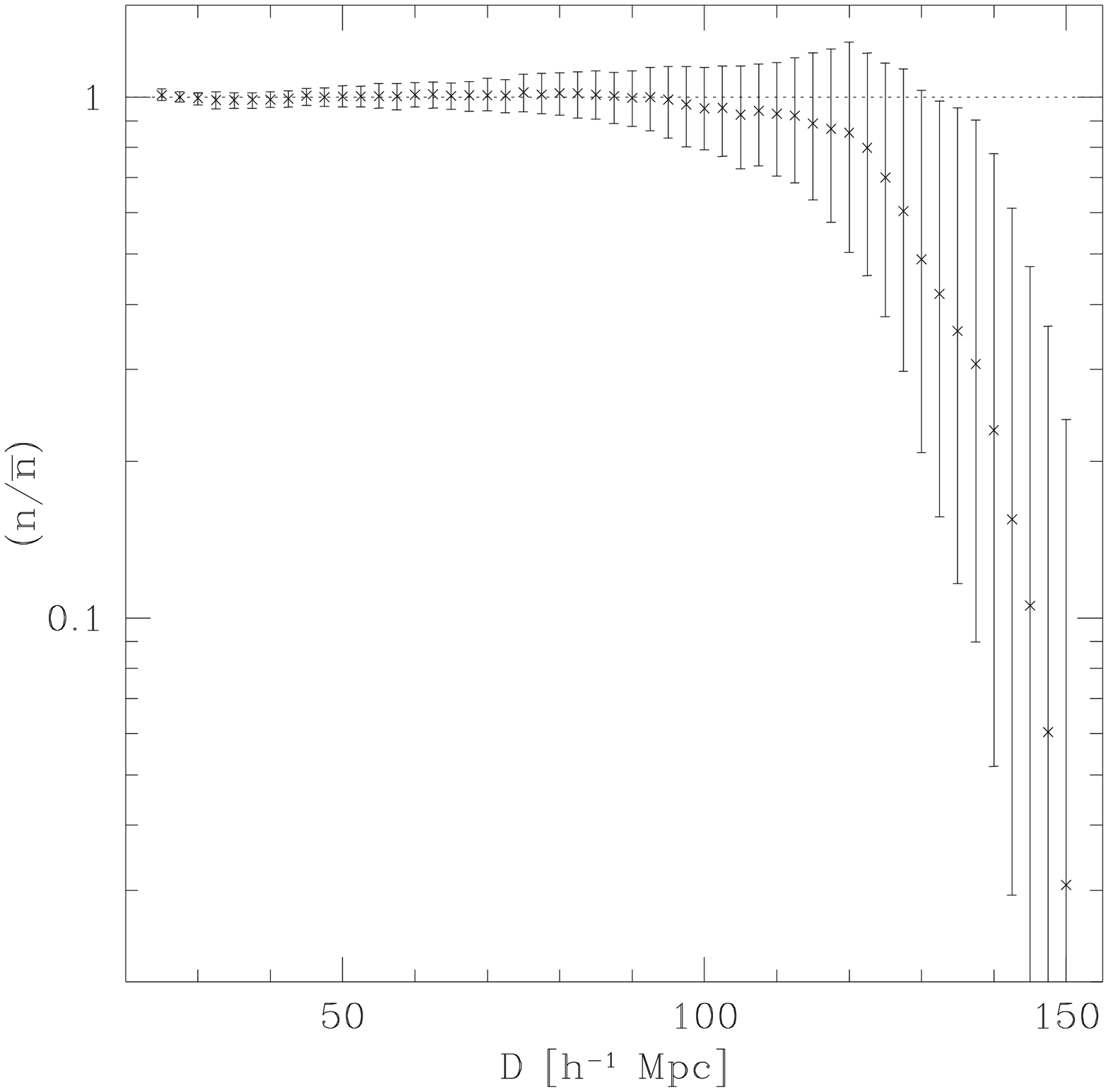}
\end{figure}
\end{document}